 reprint,
 amsmath,amssymb,
 aps,
 showpacs, showkeys, nofootinbib, longbibliography
]{revtex4-1}

\documentclass[aps,preprint,showpacs,showkeys,nofootinbib,amssymb,longbibliography]{revtex4-1} 

\usepackage{amsmath}  
\usepackage{amsfonts} 

\usepackage{url}
\usepackage{graphicx}
\usepackage{dcolumn}
\usepackage{bm}
\usepackage{hyperref}

\begin{document}

\title{Why the Tsirelson Bound?\\ Bub's Question and Fuchs' Desideratum}

\author{W.M. Stuckey}
\email{stuckeym@etown.edu}
\affiliation{
 Department of Physics\\ Elizabethtown College\\Elizabethtown, PA 17022\\
}%
\author{Michael Silberstein}
\affiliation{
 Department of Philosophy\\ Elizabethtown College\\Elizabethtown, PA 17022\\
}%
\affiliation{
Department of Philosophy\\ University of Maryland\\ College Park, MD 20742\\
}%
\author{Timothy McDevitt}
\affiliation{
Department of Mathematical Sciences\\ Elizabethtown College\\Elizabethtown, PA 17022\\
}%
\author{Ian Kohler}
\affiliation{
 Department of Engineering\\ Elizabethtown College\\Elizabethtown, PA 17022\\
}%

\date{\today}

\begin{abstract}
To answer Wheeler's question ``Why the quantum?'' via quantum information theory according to Bub, one must explain both why the world is quantum rather than classical and why the world is quantum rather than superquantum, i.e., ``Why the Tsirelson bound?'' We show that the quantum correlations and quantum states corresponding to the Bell basis states, which uniquely produce the Tsirelson bound for the Clauser-Horne-Shimony-Holt quantity, can be derived from conservation per no preferred reference frame (NPRF). A reference frame in this context is defined by a measurement configuration, just as with the light postulate of special relativity. We therefore argue that the Tsirelson bound is ultimately based on NPRF just as the postulates of special relativity. This constraint-based/principle answer to Bub's question addresses Fuchs' desideratum that we ``take the structure of quantum theory and change it from this very overt mathematical speak ... into something like [special relativity].'' Thus, the answer to Bub's question per Fuchs' desideratum is, ``the Tsirelson bound obtains due to conservation per no preferred reference frame.''
 
\end{abstract}

\pacs{03.65.−w, 03.67.−a}
\keywords{Tsirelson bound, Bell-CHSH inequality, superquantum correlations, quantum information theory}
\maketitle


\section{\label{intro}Introduction}
Wheeler's opening statement in his 1986 paper, ``How Come the Quantum?'' holds as true today as it did then \cite{wheeler1986}
\begin{quote}
The necessity of the quantum in the construction of existence: out of what deeper requirement does it arise? Behind it all is surely an idea so simple, so beautiful, so compelling that when -- in a decade, a century, or a millennium -- we grasp it, we will all say to each other, how could it have been otherwise? How could we have been so stupid for so long?
\end{quote}
The problem is, as Hardy points out, ``The standard axioms of [quantum theory] are rather ad hoc. Where does this structure come from?''\citep{hardy2016}. Concerning quantum mechanics, Fuchs writes \cite[p. 285]{fuchs1}
\begin{quote}
Compare that to one of our other great physical theories, special relativity. One could make the statement of it in terms of some very crisp and clear physical principles: The speed of light is constant in all inertial frames, and the laws of physics are the same in all inertial frames. And it struck me that if we couldn't take the structure of quantum theory and change it from this very overt mathematical speak -- something that didn't look to have much physical content at all, in a way that anyone could identify with some kind of physical principle -- if we couldn't turn that into something like this, then the debate would go on forever and ever. And it seemed like a worthwhile exercise to try to reduce the mathematical structure of quantum mechanics to some crisp physical statements. 
\end{quote}
Special relativity is a principle theory, i.e., its postulates are constraints, so quantum information theory (QIT) seeks ``the \textit{reconstruction} of quantum theory'' via a constraint-based/principle approach \citep{chiribella1} in answering Wheeler's ``Really Big Question,'' ``Why the quantum?'' \citep{wheeler1989,wheeler2004}. Indeed, QIT has produced several different sets of axioms, postulates, and ``physical requirements'' in terms of quantum information (five noted by Fuchs, for example \citep{fuchs1}) which all reproduce quantum theory. Bub has successfully recast Wheeler's question as, ``why is the world quantum and not classical, and why is it quantum rather than superquantum, i.e., why the Tsirelson bound for quantum correlations?'' \citep{bub2004,bub2012,bubbook}. 

That is, classical correlations produce a Clauser-Horne-Shimony-Holt (CHSH) quantity \citep{CHSH} between $-2$ and 2 (the Bell inequality \citep{bell}), quantum correlations produce a CHSH quantity between $-2\sqrt{2}$ and $2\sqrt{2}$ (the Tsirelson bound \citep{cirelson1980}), and superquantum correlations produce a maximum CHSH quantity of 4 (the Popescu-Rohrlich (PR) correlations \citep{PR1994}). Classical and quantum correlations exist in Nature, but superquantum correlations have not been observed. All three correlations satisfy relativistic causality (the no-signaling condition \citep{buhrman}), so  ``Why the quantum?'' meaning ``Why the quantum correlations?'' requires that we answer Bub's question, ``Why the Tsirelson bound?'' 

An interesting information-theoretic derivation of the Tsirelson bound has been produced via ``information causality'' \citep{pawlowski2009}, so it would seem QIT is making great strides in both reconstructing quantum theory and recasting and answering Wheeler's question. Bub writes, ``It's a significant sea change in the foundations of physics that information-theoretic principles of this sort are investigated as possible constraints on physical processes'' \citep[p. 183]{bubbook}. 

Despite all the success of QIT, the community does not find any of the reconstructions compelling. Cuffaro, for example, argues that information causality needs to be justified in some physical sense \citep{cuffaro1}. And, as Hardy states, ``When I started on this, what I wanted to see was two or so obvious, compelling axioms that would give you quantum theory and which no one would argue with'' \citep{Wired}. Fuchs quotes Wheeler, ``If one really understood the central point and its necessity in the construction of the world, one ought to state it in one clear, simple sentence'' \citep[p. 302]{fuchs1}. Asked if he had such a sentence, Fuchs responded, ``No, that's my big failure at this point'' \citep[p. 302]{fuchs1}.

Herein, we propose a constraint-based answer to QIT's version of ``Why the quantum?'' that we can state in ``one clear, simple sentence''
\begin{quote}
The Tsirelson bound obtains because of conservation per no preferred reference frame.
\end{quote}
Assuming the reader is willing to suspend their anthropocentric bias against constraint-based explanation \cite{silber2019,ourbook}, Section \ref{example} shows how the phenomona described by two Bell basis states, the spin-$\frac{1}{2}$ singlet state $\left(\frac{1}{\sqrt{2}} \left(\mid +1-1 \rangle - \mid -1+1 \rangle \right)\right)$ and the spin-1 `Mermin photon state' $\left(\frac{1}{\sqrt{2}} \left(\mid +1+1 \rangle + \mid -1-1 \rangle \right)\right)$, satisfy conservation of angular momentum \citep{unnik2005} per no preferred reference frame (NPRF)\footnote{We cite and use Unnikrishnan's result for the spin singlet state, but his analysis is for spin angular momentum only, which he generalizes to higher spin states, while our constraint is generalized for the conservation of anything represented by the Bell basis states. He also makes no mention of the Tsirelson bound, no preferred reference frame, special relativity, or superquantum correlations, so our analysis is a decidedly different application of this conceptual example.}. The term ``reference frame'' has many meanings in physics related to microscopic and macroscopic phenomena, Galilean versus Lorentz transformations, relatively moving observers, etc. Here, a measurement configuration constitutes a reference frame, as with the light postulate of special relativity. This constraint, conservation per NPRF, reproduces the quantum correlation function for the Bell-basis-states phenomena whence the Tsirelson bound. We then show how the quantum states themselves follow from this constraint and NPRF proper. Thus, violations of the Bell inequality up to the Tsirelson bound follow from quantum correlations obtained from the conservation of angular momentum per NPRF\footnote{It is possible for subsets of an entangled collection of particles to violate the Tsirelson bound \citep{cabello}, but of course these would not necessarily satisfy conservation of angular momentum and do not violate the Tsirelson inequality.}. It is then easy to generalize this constraint to conservation per NPRF for phenomena described by any of the Bell basis states. Since the quantum correlations and quantum states associated with the Bell-basis-states phenomena both follow from the constraint and NPRF, and the Bell basis states uniquely establish the Tsirelson bound \citep{cirelson1980,landau1987,khalfin1992}, the Tsirelson bound is ultimately grounded in NPRF, just as special relativity.

\begin{figure}[h]
\begin{center}
\includegraphics [height = 40mm]{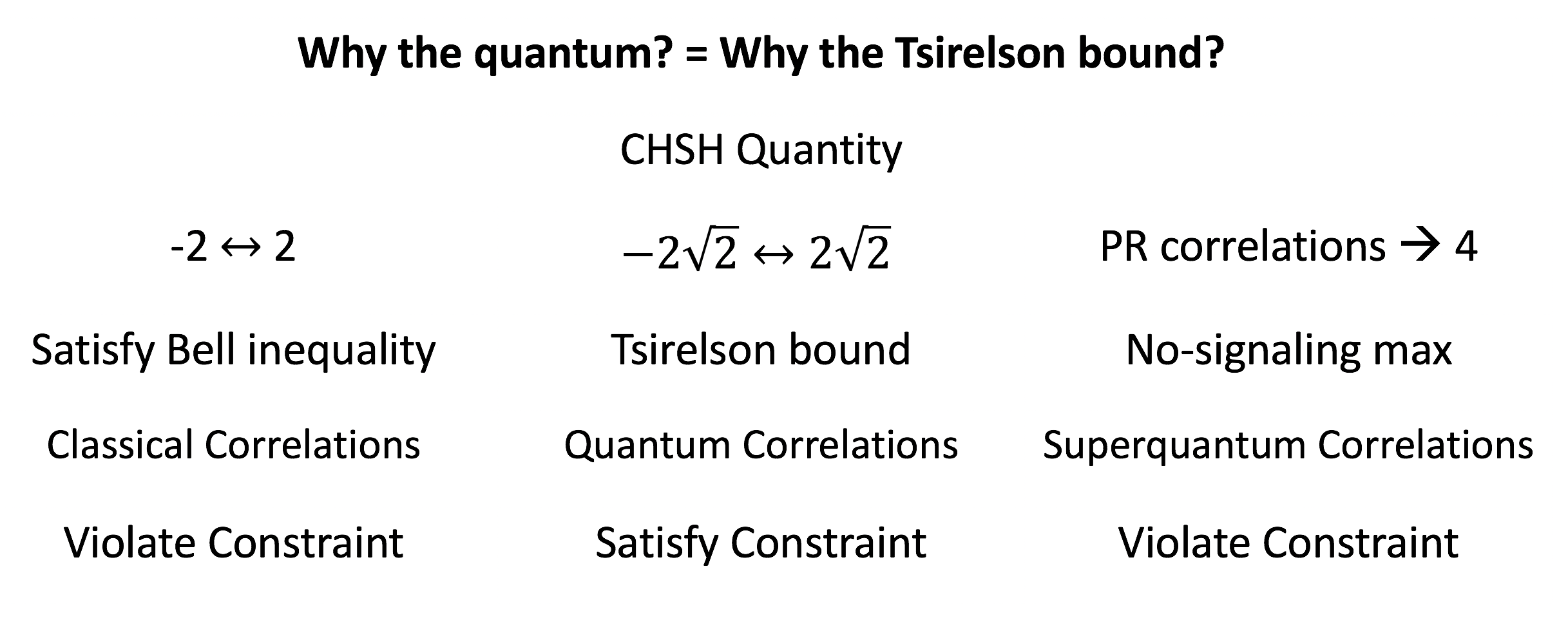}  \caption{Summary of the result. The ``constraint'' is conservation per no preferred reference frame.} \label{Summary}
\end{center}
\end{figure}

Finally, using the Popescu-Rohrlich (PR) correlations, we show how superquantum correlations that satisfy the no-signaling condition and exceed the Tsirelson bound can violate conservation per NPRF. Thus, this result shows explicitly how the quantum correlations responsible for the Tsirelson bound satisfy conservation per NPRF while both classical and superquantum correlations can violate this constraint (Figure \ref{Summary}).

\pagebreak
Our explanation of the Tsirelson bound does not require a map from 3N-dimensional Hilbert space to some `causal influence' in spacetime, e.g., a dynamical interpretation \textit{a la} Bohmian mechanics, and it does not require hidden variables (e.g., \citep{ghirardi}). That is, the $\pm 1$ outcomes for the spin-$\frac{1}{2}$ singlet state (dropping factors of $\frac{\hbar}{2}$) or the spin-1 `Mermin photon' state \citep{mermin1} (dropping factors of $\hbar$) and the angle between Stern-Gerlach (SG) magnets or polarizers are all that appear in the quantum correlations uniquely producing the Tsirelson bound and they are all that is required for our constraint-based explanation\footnote{In this analysis, we are merely pointing out how certain empirical and mathematical facts associated with the Tsirelson bound lend themselves to a parallel with a principle approach to special relativity, precisely per Fuchs' desideratum. Therefore, our analysis does not and cannot rule out any particular interpretation of quantum mechanics. This includes any particular interpretation of quantum probability.}. 

Thus, we see explicitly in this result how quantum mechanics conforms statistically to a conservation principle without need of a `causal influence' or hidden variables acting on a trial-by-trial basis to account for that conservation. There are many attempts to add such classical mechanisms, but they are superfluous as far as the physics is concerned. The light postulate of special relativity is a good analogy for our proposed constraint. 

In special relativity, Alice is moving at velocity $\vec{V}_a$ relative to a light source and measures the speed of light from that source to be \textit{c}. Bob is moving at velocity $\vec{V}_b$ relative to that same light source and measures the speed of light from that source to be \textit{c}. Here ``reference frame'' refers to the relative motion of the observer and source, so all observers who share the same relative velocity with respect to the source occupy the same reference frame. NPRF in this context thus means all measurements produce the same outcome \textit{c}. 

As a consequence of this constraint, Alice says Bob must correct his length and time measurements per length contraction and time dilation, while Bob says the same thing about Alice's measurements. This apparent contradiction is then reconciled per NPRF via the relativity of simultaneity. That is, Alice and Bob each partition spacetime per their own equivalence relations (per their own reference frames), so that equivalence classes are their own surfaces of simultaneity. If Alice's equivalence relation over the spacetime events yields the ``true'' partition of spacetime, then Bob must correct his lengths and times per length contraction and time dilation. Of course, the relativity of simultaneity says that Bob's equivalence relation is as valid as Alice's per NPRF. As a consequence of this constraint and NPRF proper (giving the relativity postulate of special relativity), special relativity is a constraint-based/principle theory.

\begin{figure}[h]
\begin{center}
\includegraphics [height = 40mm]{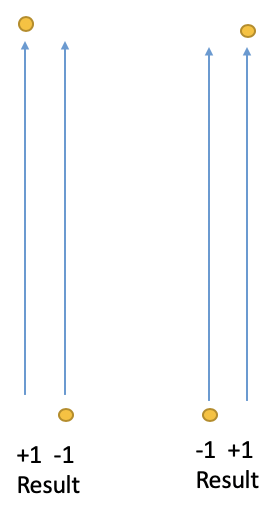}  \caption{Outcomes (yellow dots) in the same reference frame, i.e., outcomes for the same measurement (blue arrows represent SG magnet orientations), for the spin-$\frac{1}{2}$ singlet state explicitly conserve angular momentum. Note: Herein ``spin singlet state'' means ``spin-$\frac{1}{2}$ singlet state.''} \label{LikeOrient}
\end{center}
\end{figure}

In quantum mechanics, Alice orients her SG magnet at $\alpha$ relative to a source of spin entangled particles and measures +1 or --1 ($\frac{\hbar}{2}$). Bob orients his SG magnet at $\beta$ relative to that same source of spin entangled particles and measures +1 or --1 ($\frac{\hbar}{2}$). NPRF in this context means all measurements produce the same outcome +1 or --1 ($\frac{\hbar}{2}$). As a consequence of this constraint, we can only conserve angular momentum on average between different reference frames, i.e., it cannot be conserved on a trial-by-trial basis unless the SG magnets or polarizers are co-aligned (Figure \ref{LikeOrient}). NPRF reconciles this conflict via the ``relativity of data partition.'' That is, Alice and Bob each partition the data per their own equivalence relations (per their own reference frames), so that equivalence classes are their own $+1$ and $-1$ data events. If Alice's equivalence relation over the data events yields the ``true'' partition of the data, then Bob must correct (average) his results per average-only conservation. Of course, NPRF says that Bob's equivalence relation is as valid as Alice's, which we might call the ``relativity of data partition'' (Figure \ref{SRvQM}). 

\begin{figure}[h]
\begin{center}
\includegraphics [height = 28mm]{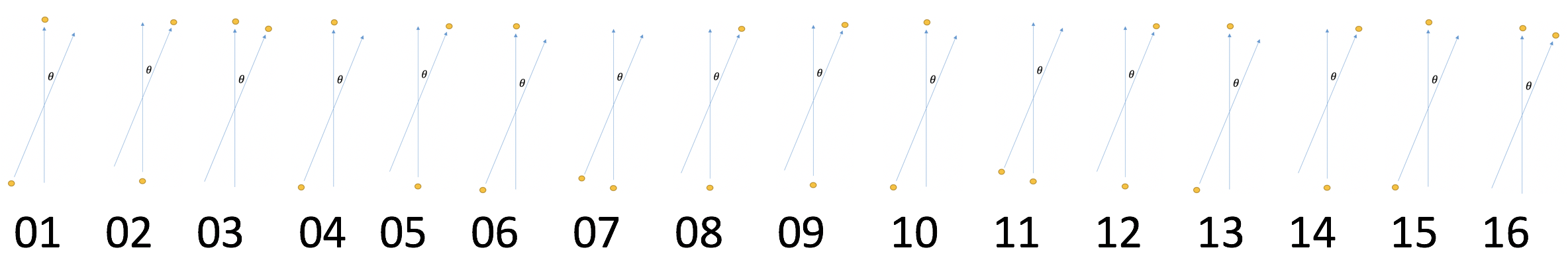}  \caption{A spatiotemporal ensemble of 16 experimental trials for the spin singlet state. Angular momentum is not conserved in any given trial, because there are two different measurements being made, i.e., outcomes are in two different reference frames, but it is conserved on average for all 16 trials. It is impossible for angular momentum to be conserved explicitly in each trial since the measurement outcomes are binary (quantum) with values of +1 (up) or --1 (down) $\left(\frac{\hbar}{2}\right)$ per no preferred reference frame. The conservation principle at work here assumes Alice and Bob's measured values of angular momentum are not mere components of some hidden angular momentum, e.g., oppositely oriented $\vec{S}_A$ and $\vec{S}_B$ so that Alice and Bob's $\pm 1$ results are always components of those hidden $\vec{S}_A$ and $\vec{S}_B$ with variable magnitudes. That is, the measured values of angular momentum \textit{are} the angular momenta contributing to this conservation principle.} \label{4Dpattern}
\end{center}
\end{figure}

\begin{figure}[h]
\begin{center}
\includegraphics [height = 25mm]{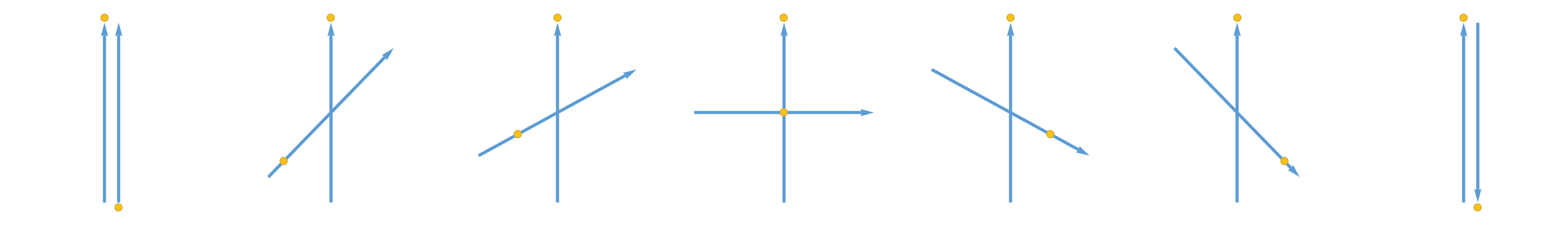}  \caption{Reading from left to right, as Bob rotates his SG magnets relative to Alice's SG magnets for her +1 outcome, the average value of his outcome varies from --1 (totally down, arrow bottom) to 0 to +1 (totally up, arrow tip). This obtains per conservation of angular momentum on average in accord with no preferred reference frame. Bob can say exactly the same about Alice's outcomes as she rotates her SG magnets relative to his SG magnets for his +1 outcome. That is, their outcomes can only satisfy conservation of angular momentum on average in different reference frames, because they only measure $\pm 1\left(\frac{\hbar}{2}\right)$, never a fractional result. Thus, just as with the light postulate of special relativity, we see that no preferred reference frame requires quantum ($\pm 1$) outcomes for all measurements and that leads to a constraint-based/principle answer to Bub's question.} \label{AvgView}
\end{center}
\end{figure}

\begin{figure}[h]
\begin{center}
\includegraphics [height = 40mm]{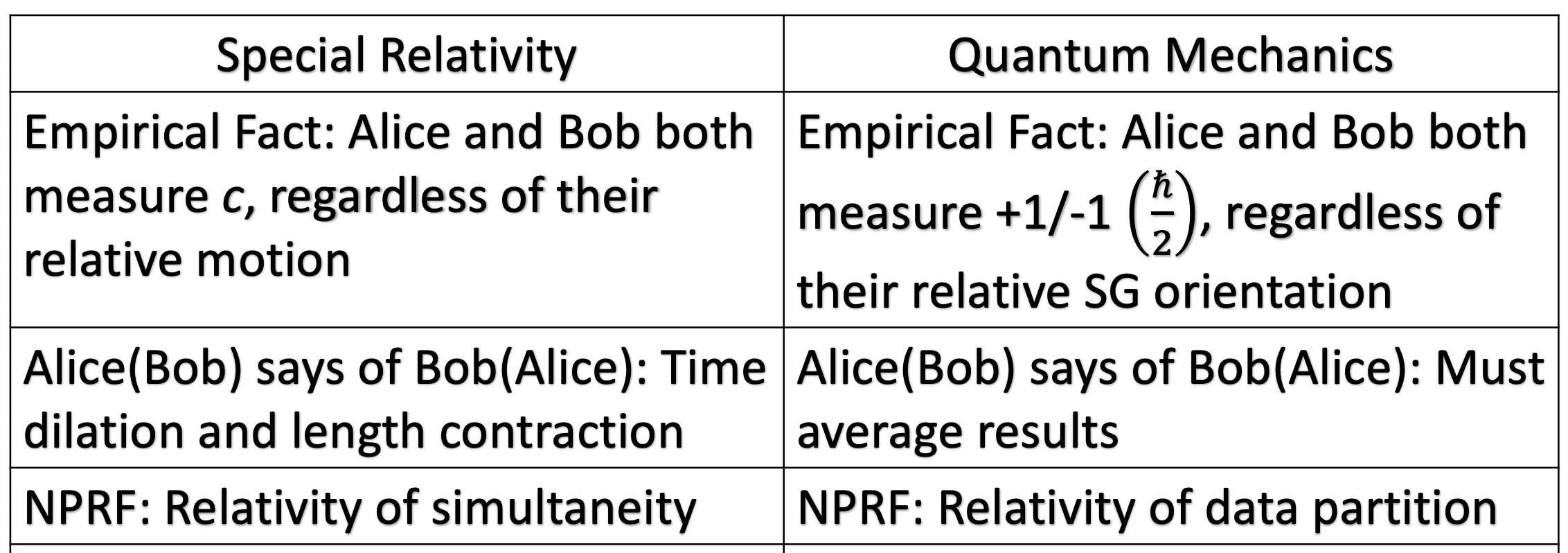}  \caption{Comparing special relativity with quantum mechanics according to no preferred reference frame (NPRF). Because Alice and Bob both measure the same speed of light \textit{c} regardless of their relative motion per NPRF, Alice(Bob) may claim that Bob's(Alice's) length and time measurements are erroneous and need to be corrected (length contraction and time dilation). Likewise, because Alice and Bob both measure the same values for spin angular momentum $\pm 1$ $\left(\frac{\hbar}{2}\right)$ regardless of their relative SG magnet orientation per NPRF, Alice(Bob) may claim that Bob's(Alice's) individual $\pm 1$ values are erroneous and need to be corrected (averaged, Figures \ref{4Dpattern} \& \ref{AvgView}). In both cases, NPRF resolves the mystery it creates. In special relativity, the apparently inconsistent results can be reconciled via the relativity of simultaneity. That is, Alice and Bob each partition spacetime per their own equivalence relations (per their own reference frames), so that equivalence classes are their own surfaces of simultaneity and these partitions are equally valid per NPRF. This is completely analogous to quantum mechanics, where the apparently inconsistent results per the Bell spin states arising because of NPRF can be reconciled by NPRF via the ``relativity of data partition.'' That is, Alice and Bob each partition the data per their own equivalence relations (per their own reference frames), so that equivalence classes are their own $+1$ and $-1$ data events and these partitions are equally valid.} \label{SRvQM}
\end{center}
\end{figure}

As we will show, this constraint plus NPRF proper (giving $P_{+-} = P_{-+}$) then produce the quantum state, i.e., probability for each possible measurement outcome. This is quite unlike classical physics (Figures \ref{4Dpattern} \& \ref{AvgView}); in fact it is what uniquely distinguishes the quantum joint distribution from its classical counterpart \citep{garg}. Thus, NPRF here leads to quantum outcomes ($\pm 1$ only) and `average-only' conservation (Figure \ref{SRvQM}), a constraint-based/principle answer to Bub's question. So, our answer to Bub's question satisfying Fuchs' desideratum is
\begin{quote}
The Tsirelson bound obtains because of conservation per no preferred reference frame.
\end{quote}

We do acknowledge that our explanation of the Tsirelson bound lies outside the QIT enterprise devoted to explaining quantum probability theory via information-theoretic principles over all possible probability structures \citep{hardy2011,hardy2016,fuchs1}. In the parlance of metaphysics, we have not ruled out some ``possible world'' in which superquantum correlations exist, we have only ruled them out on empirical grounds for our world in accord with physics. However, the result is not without relevance for QIT. From an information-theoretic perspective, what is actually being conserved in this fashion is the fundamental unit of binary information, e.g., on-off(ness) or on-on(ness) or yes-no(ness), etc. 

This is not conservation of information \textit{a la} information causality; our constraint does not deal with signals sent between Alice and Bob, but with the spatiotemporal correlations in their measurement settings and results. Accordingly, we concur with the QIT approach to view quantum mechanics in terms of spacetime constraints on par with a principle theory like special relativity (Figure \ref{SRvQM}), rather than dynamical laws or processes. Therefore, we feel this constraint-based explanation of the Tsirelson bound does contribute to the desideratum of QIT. Again, we emphasize that this is a constraint-based explanation of the Tsirelson bound per physics, so it cannot satisfy any grander extra-physical expectations, i.e., pure information-theoretic constraints, of some in QIT.

\section{\label{example}The Tsirelson Bound from a Conservation Principle}
We assume at this point the reader is prepared to consider a constraint-based/principle \textit{explanation} \cite{silber2019,ourbook} of the phenomena responsible for the Tsirelson bound. We will present the physics that is germane to our argument though much of it is doubtless familiar to the reader. The deeper point is to look at the physics via constraints (Figures \ref{4Dpattern} \& \ref{AvgView}), so as to remove the mystery of entanglement created by dynamical bias (Section \ref{disc}). While the following analysis itself does not require it, we urge the reader to take seriously the possibility that quantum entanglement is explained not by some dynamical/causal process in Hilbert space or spacetime, but by constraints on events in spacetime \textit{a la} a principle theory. 

We start with a specific case, viz., conservation of angular momentum per NPRF, before generalizing that to conservation per NPRF. More specifically, we consider the spin singlet state (total anti-correlation) and the `Mermin state' \citep{mermin1} for photons (total correlation) since they are examples of two Bell basis states with obvious physical meaning. After our presentation of this transparent conservation principle, we will show how it and NPRF proper give the quantum states. We can then show how conservation of angular momentum per NPRF generalizes to conservation per NPRF for Bell-basis-states phenomena.

In principle, the creation of an entangled state due to conservation of angular momentum is not difficult to imagine, e.g., the dissociation of a spin-zero diatomic molecule \citep{bohm} or the decay of a neutral pi meson into an electron-positron pair \citep{larosa}. In reality, creating a spin singlet state or the Mermin photon state in a controlled experimental situation is nontrivial \citep{hensen,dehlinger}, i.e., the preparation fragment of the circuit would contain many operations and wires, so we will suppress the preparation into a spatially localized region which we will call ``the source'' per convention (Hardy calls this an ``equivalence class of preparations'' \citep{hardy2011}). The spin singlet state (total angular momentum equals zero \citep[p. 29-30]{renes}, \citep{MerminChallenge}) is $\frac{1}{\sqrt{2}} \left(\mid ud \rangle - \mid du \rangle \right)$ where \textit{u}/\textit{d} means the outcome is displaced upwards/downwards relative to the north-south pole alignment of the SG magnets. This state obtains due to conservation of angular momentum at the source as represented by momentum exchange in the spatial plane orthogonal to the source collimation (binary information is ``up or down'' transverse). 
The Mermin state for photons (a Bell basis state in the triplet space with total angular momentum of 2 in units of $\hbar = 1$ \citep[p. 29-30]{renes}, \citep{MerminChallenge}) is $\frac{1}{\sqrt{2}} \left(\mid VV \rangle + \mid HH \rangle \right)$ where \textit{V} (``vertical'') means the there is an outcome (photon detection) behind one of the co-aligned polarizers and \textit{H} (``horizontal'') means there is no outcome behind one of the co-aligned polarizers. This state obtains due to conservation of angular momentum at the source as represented by momentum exchange along the source collimation (binary information is ``yes or no'' longitudinal). At this point we will focus the discussion on the spin singlet state for total anti-correlation, since everything said of that state can be easily transferred to the Mermin photon state. 

\begin{figure}
\begin{center}
\includegraphics [height = 40mm]{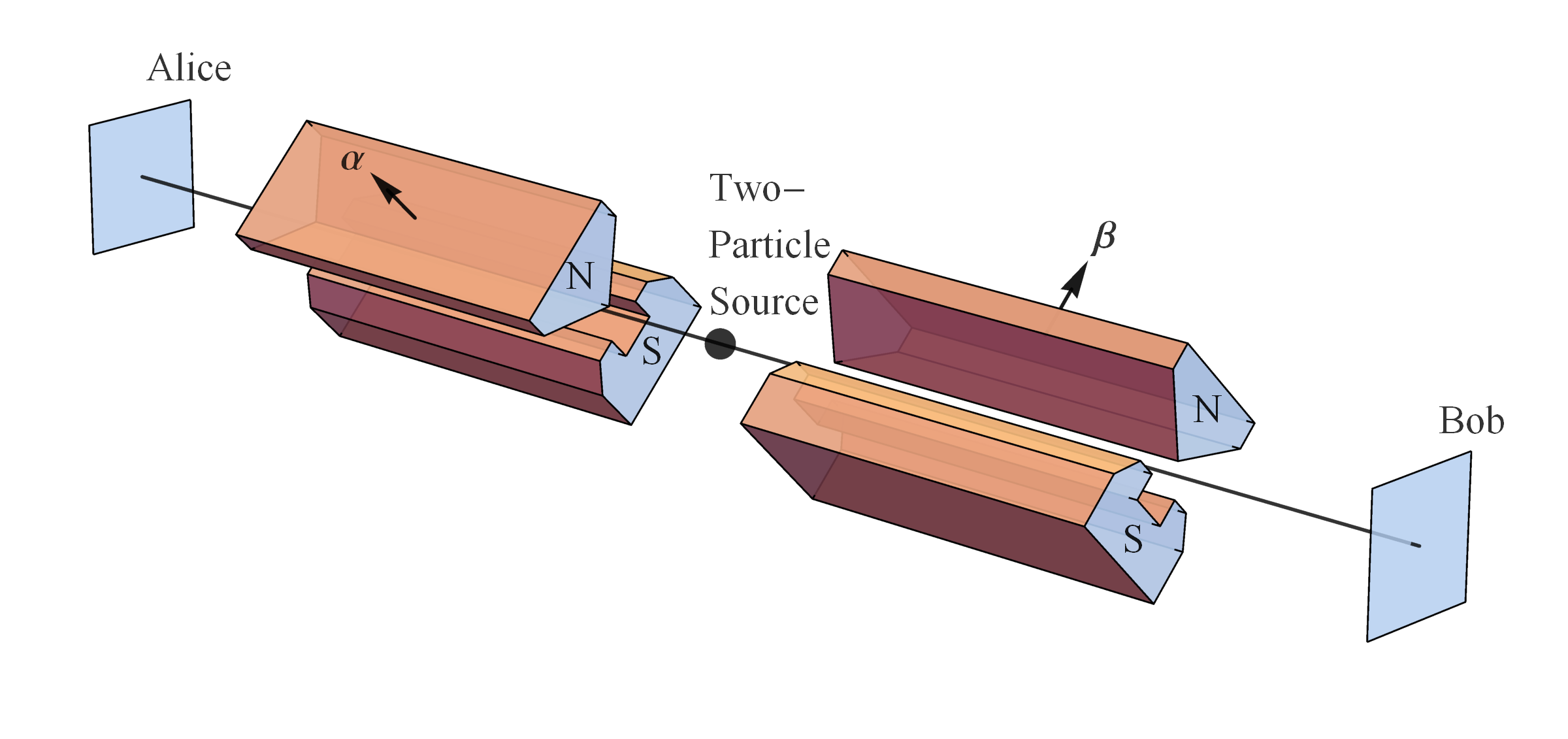}  \caption{Alice and Bob making spin measurements with their Stern-Gerlach (SG) magnets and detectors.} \label{EPRB}
\end{center}
\end{figure}

We wire the preparation to a pair of transformations then wire those to a pair of results, i.e., we orient the emission directions of the source towards SG magnets and detectors (Figure \ref{EPRB}).
\begin{figure}
\begin{center}
\includegraphics [height = 40mm]{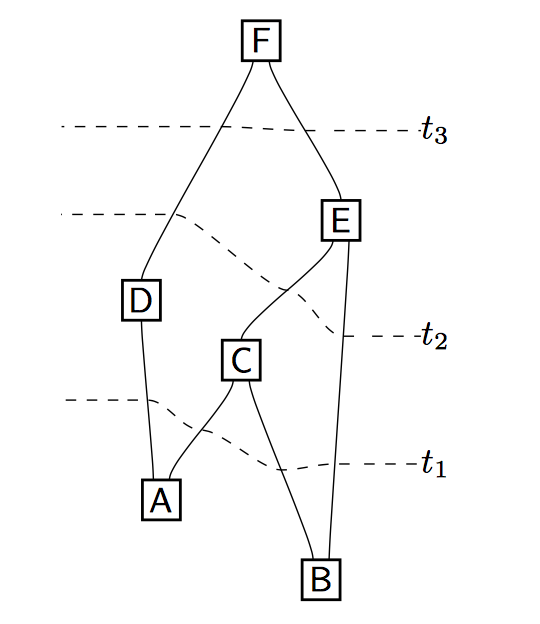}  \caption{``It is always possible to provide a complete foliation for a circuit'' \citep[p. 20]{hardy2011}. Reproduced with permission from the author.} \label{hardycircuit}
\end{center}
\end{figure}
The circuits of QIT represent a series of operations in space and time (Figure \ref{hardycircuit}) in accord with our attempt to explain the spatiotemporal ensemble of spatiotemporal experimental trials for the spin singlet and Mermin photon states (Figure \ref{4Dpattern}). In this way we move the discussion from time-evolved state vectors in 3N-dimensional Hilbert space to the computation of probabilities for circuits in spacetime using the Feynman path integral. We can obtain the following probability amplitudes for the various correlated outcomes of the circuits associated with our spin singlet state using the path integral method \citep{sinha,wharton1} 
\begin{equation}
A_{ud}= -A_{du} =\frac{1}{2 \sqrt{2}}\left(e^{i\alpha} + e^{i\beta}\right)
\end{equation}
\begin{equation}
A_{uu}= -A_{dd} = \frac{1}{2 \sqrt{2}}\left(e^{i\alpha} - e^{i\beta}\right)
\end{equation}
where $\alpha$ and $\beta$ are the SG magnet orientation angles in space (Figure \ref{EPRB}). The corresponding probabilities are then
\begin{equation}
P_{ud} = \left| A_{ud} \right|^2 = \left| A_{du} \right|^2 = P_{du} = \frac{1}{2} \mbox{cos}^2 \left(\frac{\alpha - \beta}{2}\right) \label{probabilityud}
\end{equation}
and
\begin{equation}
P_{uu} = \left| A_{uu} \right|^2 = \left| A_{dd} \right|^2 = P_{dd} = \frac{1}{2} \mbox{sin}^2 \left(\frac{\alpha - \beta}{2}\right) \label{probabilityuu}
\end{equation}
Eqs. (\ref{probabilityud}) \& (\ref{probabilityuu}) constitute the ``quantum state'' for the spin singlet state. We will derive these from our constraint and NPRF proper below. Likewise, the probabilities for the Mermin photon state are
\begin{equation}
P_{VV} = P_{HH} = \frac{1}{2} \mbox{cos}^2 \left(\alpha - \beta \right) \label{probabilityVV}
\end{equation}
and
\begin{equation}
P_{VH} = P_{HV} = \frac{1}{2} \mbox{sin}^2 \left(\alpha - \beta \right) \label{probabilityVH}
\end{equation}
Equations (\ref{probabilityVV}) \& (\ref{probabilityVH}) constitute the quantum state for the Mermin photon state. We will also derive these from our constraint and NPRF proper below. Each probability for a particular circuit is spatiotemporal in that it is the probability for a source emission event with its corresponding orientations of the two SG magnets ($\alpha$ and $\beta$) and two outcomes (\textit{uu}, \textit{dd}, \textit{ud}, or \textit{du}), as represented by each trial in Figure \ref{4Dpattern}. Let us investigate what Alice and Bob discover about this preparation in the various spatiotemporal classical contexts of their transformations and results. 

Alice's detector responds up and down with equal frequency regardless of the orientation $\alpha$ of her SG magnet, i.e., she obtains the same outcome regardless of her measurement per NPRF. Bob observes the same regarding his SG magnet orientation $\beta$. Thus, the source is rotationally invariant in the spatial plane orthogonal to the source collimation in this spacetime context per NPRF \cite{MerminChallenge}. When Bob and Alice compare their outcomes, they find that their outcomes are perfectly anti-correlated (\textit{ud} and \textit{du} with equal frequency) when $\alpha - \beta = 0$, i.e., when making the same measurement (meaning they are in the same reference frame). This is consistent with conservation of angular momentum per classical mechanics between the pair of detection events (this fact defines the state). The degree of that anti-correlation diminishes as $\alpha - \beta \rightarrow \frac{\pi}{2}$ until it is equal to the degree of correlation (\textit{uu} and \textit{dd}) when their SG magnets are at right angles to each other. In other words, whenever the SG magnets are orthogonal to each other anti-correlated and correlated outcomes occur with equal frequency, i.e., conservation of angular momentum in one direction is independent of the angular momentum changes in any orthogonal direction. Thus, we would not expect to see more correlation or more anti-correlation based on conservation of angular momentum for transverse results in the spatial plane orthogonal to the source collimation when the SG magnets are orthogonal to each other. Notice that once their SG magnets are not co-aligned, the conservation between them obtains in a purely statistical sense, since we can no longer have explicit cancellation of Alice and Bob's measured values (Figure \ref{AvgView}). 

As we continue to increase the angle $\alpha - \beta$ beyond $\frac{\pi}{2}$ the anti-correlations continue to diminish until we have totally correlated outcomes when the SG magnets are anti-aligned. This is also consistent with conservation of angular momentum, since the totally correlated results when the SG magnets are anti-aligned represent momentum exchanges in opposite directions in the spatial plane orthogonal to the source collimation just as when the SG magnets are aligned, it is now simply the case that what Alice calls up, Bob calls down and vice-versa. 

The counterpart for the Mermin photon state is simply that angular momentum conservation is evidenced by \textit{VV} or \textit{HH} outcomes for co-aligned polarizers (again, this fact defines the state), i.e., when making the same measurement (meaning they are in the same reference frame). When the polarizers are at right angles you have only VH and HV outcomes, which is still totally consistent with conservation of angular momentum as `not \textit{H}' implies \textit{V} and vice-versa \citep{dehlinger}. In other words, a polarizer does not have a `north-south' distinction (longitudinal rather than transverse momentum exchange). In particular, having rotated either or both polarizers by $\pi$ one should obtain precisely \textit{VV} or \textit{HH} outcomes again. 

One might then ask (rhetorically for this audience) whether or not it is possible to explain the conservation of angular momentum per NPRF for this spin singlet circuit using classical probability theory. Of course, that would be a hidden variable theory amenable to counterfactual definiteness (Mermin's ``instruction sets'' \citep{mermin1,mermin2,MerminChallenge}) on a trial-by-trial basis. This possibility is explored using correlations where the probability of outcomes \textit{i} and \textit{j} for settings \textit{a} and \textit{b} is written $p(i,j \mid a,b)$ \citep{cuffaro1}. We start with the fact that Alice's outcomes are not influenced by Bob's settings and vice-versa
\begin{equation}
\begin{split}
p(A \mid a\phantom{\prime},b\phantom{\prime}) &= p(A \mid a\phantom{\prime}, b^\prime)\\
p(A \mid a^\prime,b\phantom{\prime}) &= p(A \mid a^\prime, b^\prime)\\
p(B \mid a\phantom{\prime},b\phantom{\prime}) &= p(B \mid a^\prime, b\phantom{\prime})\\
p(B \mid a\phantom{\prime},b^\prime) &= p(B \mid a^\prime, b^\prime ) \label{nosig}
\end{split}
\end{equation}
This is the no-signaling condition alluded to earlier. Next, we write the average of outcomes $i \cdot j$ for settings \textit{a} and \textit{b} as
\begin{equation}
\langle a,b \rangle = \sum (i \cdot j) \cdot p(i,j \mid a,b)  \label{average}
\end{equation}
and construct the Clauser-Horne-Shimony-Holt (CHSH) quantity
\begin{equation}
\langle a,b \rangle + \langle a,b^\prime \rangle + \langle a^\prime,b \rangle - \langle a^\prime,b^\prime \rangle \label{CHSH1}
\end{equation}
If one attempted to instantiate the momentum exchanges of the spin singlet circuit using instruction sets/counterfactual definiteness/hidden variables per classical probability theory in accord with no-signaling, Eq. (\ref{CHSH1}) would give a value between 2 and --2 (CHSH version of Bell's inequality \citep{bell}). However, Eqs. (\ref{probabilityud}) and (\ref{probabilityuu}) per quantum mechanics put into Eqs. (\ref{average}) and (\ref{CHSH1}) give 
\begin{equation}
-\cos(a - b) -\cos(a - b^\prime) -\cos(a^\prime - b) +\cos(a^\prime - b^\prime) \label{CHSHspin}
\end{equation}
Choosing $a = \pi/4$, $a^\prime = -\pi/4$, $b = 0$, and $b^\prime = \pi/2$ minimizes Eq. (\ref{CHSHspin}) at $-2\sqrt{2}$ (the Tsirelson bound).

Everything said here concerning angular momentum conservation per anti-correlated outcomes of the spin singlet state applies for angular momentum conservation per correlated outcomes of the Mermin photon state, which gives
\begin{equation}
\cos2(a - b) +\cos2(a - b^\prime) +\cos2(a^\prime - b) -\cos2(a^\prime - b^\prime) \label{CHSHmermin}
\end{equation}
for Eq. (\ref{CHSH1}). Using $a = \pi/8$, $a^\prime = -\pi/8$, $b = 0$, and $b^\prime = \pi/4$ maximizes Eq. (\ref{CHSHmermin}) at $2\sqrt{2}$ (the Tsirelson bound). Experiments show that the quantum results can be achieved (violating the Bell inequality), ruling out an explanation of these correlated momentum exchanges via classical probability theory. We now derive these quantum correlations using the conservation of angular momentum \citep{unnik2005} per NPRF. Using that result and NPRF proper, we will then be able to derive the quantum states and generalize the constraint to conservation per NPRF.

It is easy to see how this follows by starting with total angular momentum of zero for binary (quantum) outcomes\footnote{Again, this argument is for the spin-$\frac{1}{2}$ singlet state and we are suppressing the factor of $\frac{\hbar}{2}$. ``Binary'' entails ``quantum'' so we will stop qualifying it as such.} +1 and --1. Alice and Bob both measure +1 and --1 results with equal frequency for any SG magnet angle (NPRF) and when their angles are equal they obtain totally anti-correlated outcomes giving total angular momentum of zero (this is a defining factor). Now divide Alice's results into two subsets of +1 and --1, each occurring $\frac{N}{2}$ times when the total number of measurements is $N$ (the argument is symmetric with respect to Bob, obviously). Contrary to classical mechanics, conservation of angular momentum per NPRF does not allow us to make a definitive prediction about a particular outcome at Bob's location based on a particular corresponding outcome at Alice's location when Bob's SG magnet is rotated by $\theta$ relative to Alice's, because Bob only ever measures +1 or --1 per NPRF, i.e., no fractions\footnote{Again, this fact alone distinguishes the quantum joint distribution from its classical counterpart \citep{garg}.} (Figure \ref{4Dpattern}). 

We have two sets of data, Alice's set and Bob's set. They were collected in $N$ pairs (data events) with Bob's(Alice's) SG magnet at $\theta$ relative to Alice's(Bob's). We want to compute the correlation function for these $N$ data events which is

\begin{equation}\langle \alpha,\beta \rangle = \frac{(+1)_A(-1)_B + (+1)_A(+1)_B + (-1)_A(-1)_B + ...}{N}\end{equation}
Now partition the numerator into two equal subsets per Alice's equivalence relation, i.e., Alice's $+1$ results and Alice's $-1$ results

\begin{equation}\langle \alpha,\beta \rangle = \frac{(+1)_A(\sum \mbox{BA+})+(-1)_A(\sum \mbox{BA-})}{N}\label{correl2}\end{equation}
where $\sum \mbox{BA+}$ is the sum of all of Bob's results corresponding to Alice's result $(+1)_A$ and $\sum \mbox{BA-}$ is the sum of all of Bob's results corresponding to Alice's result $(-1)_A$. Again, we could just as well have used Bob's results $(+1)_B$ and $(-1)_B$ and obtained averages over Alice's results instead, since the situation is symmetric (NPRF). Now, rewrite Eq. (\ref{correl2}) as

\begin{equation}\langle \alpha,\beta \rangle = \frac{1}{2}(+1)_A\overline{BA+} + \frac{1}{2}(-1)_A\overline{BA-} \label{correl3} \end{equation}
with the overline denoting average. In order to obtain the quantum correlation function we need some principle that specifies $\overline{BA+}$ and $\overline{BA-}$ and we are proposing that principle is our particular form of conservation. Here is how one might argue for the principle using classical reasoning applied to the quantum exchange of momentum. 

The projection of the angular momentum vector of Alice's particle $\vec{S}_A = +1\hat{\alpha}$ along $\hat{\beta}$ is $\vec{S}_A\cdot\hat{\beta} = +\cos{\theta}$ where again $\theta$ is the angle between the unit vectors $\hat{\alpha}$ and $\hat{\beta}$. From Alice's perspective, when Bob makes the same measurement, i.e., $\beta = \alpha$, he finds the angular momentum vector of his particle is $\vec{S}_B = -1\hat{\alpha}$, so that $\vec{S}_A + \vec{S}_B = \vec{S}_{Total} = 0$. When he does not make the same measurement, he should obtain a fraction of the length of $\vec{S}_B$, i.e., $\vec{S}_B\cdot\hat{\beta} = -1\hat{\alpha}\cdot\hat{\beta} = -\cos{\theta}$ (this also follows from counterfactual spin measurements on the single-particle state \citep{boughn}). Of course per NPRF, Bob only ever obtains +1 or --1, so we posit that Bob will \textit{average} the required $-\cos{\theta}$ (Figures \ref{4Dpattern} \& \ref{AvgView}), which means

\begin{equation}\overline{BA+} = -\cos\theta\label{AGC1}\end{equation}
Likewise, for Alice's $(-1)_A$ results we have
\begin{equation}\overline{BA-} = \cos\theta\label{AGC2}\end{equation}
Putting these into Eq. (\ref{correl3}) we obtain
\begin{equation}
\langle \alpha,\beta \rangle = \frac{1}{2}(+1)_A(-\cos\theta) + \frac{1}{2}(-1)_A(\cos\theta) = -\cos\theta 
\end{equation}
which is precisely the correlation function given by quantum mechanics. This derivation of the quantum correlation function is independent of the formalism of quantum mechanics, instead it follows from a compelling and simple physical principle, the conservation of angular momentum per NPRF. Now let us use this result and derive the corresponding quantum state, i.e., Eqs. (\ref{probabilityud}) \& (\ref{probabilityuu}).

We need to find $P_{uu}$, $P_{dd}$, $P_{ud}$, and $P_{du}$ so we need four independent conditions. Normalization gives
\begin{equation}
P_{uu} + P_{ud} + P_{du} + P_{dd} = 1\label{normalization}
\end{equation}
and our correlation function
\begin{equation}
\langle \alpha,\beta \rangle = (+1)_A(+1)_BP_{uu} + (+1)_A(-1)_BP_{ud} + (-1)_A(+1)_BP_{du} + (-1)_A(-1)_BP_{dd}\label{correlFn}
\end{equation}
along with our constraint represented by Eqs. (\ref{correl3} -- \ref{AGC2}) give
\begin{equation}
P_{uu} - P_{ud} = -\frac{1}{2}\cos\theta
\end{equation}
and
\begin{equation}
P_{du} - P_{dd} = \frac{1}{2}\cos\theta
\end{equation}
Finally, NPRF gives $P_{ud} = P_{du}$, since $P_{ud}$ is Alice's up results paired with Bob's down results and $P_{du}$ is Bob's up results paired with Alice's down results. Solving these four equations for $P_{uu}$, $P_{dd}$, $P_{ud}$, and $P_{du}$ gives precisely Eqs. (\ref{probabilityud}) \& (\ref{probabilityuu}). 

Notice that since the angle between SG magnets $\alpha - \beta$ is twice the angle between Hilbert space measurement bases, this result easily generalizes to conservation per NPRF of whatever the measurement outcomes represent when unlike outcomes entail conservation. All one need do is let $\theta \rightarrow 2\theta$ in the above analysis, understanding that $\theta$ now represents the angle between Hilbert space measurement bases. The origin of the more general conservation principle is transparent in the next example where the angle between polarizers $\alpha - \beta$ equals the angle between Hilbert space measurement bases.

For the Mermin photon state, conservation of angular momentum is established by pass (designated by +1) and no pass (designated by --1) results through a polarizer. When the polarizers are co-aligned Alice and Bob get the same results, half pass and half no pass. Thus, conservation of angular momentum is established by the intensity of the electromagnetic radiation applied to binary outcomes for various polarizer orientations. Again, grouping Alice's results into +1 and --1 outcomes we see that she would expect Bob to measure $[\mbox{cos}^2\theta - \mbox{sin}^2\theta]$ at $\theta$ for her +1 results and $[\mbox{sin}^2\theta - \mbox{cos}^2\theta]$ for her --1 results. Since Bob measures the same thing as Alice for conservation of angular momentum, we posit that those are Bob's averages when his polarizer deviates from Alice's by $\theta$. Therefore, the correlation $\langle \alpha,\beta \rangle$ of results for conservation of angular momentum per Equation (\ref{correl3}) is 
\begin{equation}
\frac{1}{2}(+1)_A(\mbox{cos}^2\theta - \mbox{sin}^2\theta) + \frac{1}{2}(-1)_A(\mbox{sin}^2\theta - \mbox{cos}^2\theta) = \cos{2\theta} \label{merminconserve} 
\end{equation}
which is precisely the correlation function given by quantum mechanics. Now let us use this result and derive the corresponding quantum state, i.e., Eqs. (\ref{probabilityVV}) \& (\ref{probabilityVH}).

As before, we need to find $P_{VV}$, $P_{HH}$, $P_{VH}$, and $P_{HV}$ so we need four independent conditions. Normalization and $P_{VH} = P_{HV}$ are the same as for the spin singlet case. The correlation function
\begin{equation}
\langle \alpha,\beta \rangle = (+1)_A(+1)_BP_{VV} + (+1)_A(-1)_BP_{VH} + (-1)_A(+1)_BP_{HV} + (-1)_A(-1)_BP_{HH}\label{correlFn2}
\end{equation}
along with our constraint represented by Eq. (\ref{merminconserve}) give
\begin{equation}
P_{VV} - P_{VH} = -\frac{1}{2}(\mbox{sin}^2\theta - \mbox{cos}^2\theta)
\end{equation}
and
\begin{equation}
P_{HV} - P_{HH} = \frac{1}{2}(\mbox{sin}^2\theta - \mbox{cos}^2\theta)
\end{equation}
Solving these four equations for $P_{VV}$, $P_{HH}$, $P_{VH}$, and $P_{HV}$ gives precisely Eqs. (\ref{probabilityVV}) \& (\ref{probabilityVH}). 

Notice that since the angle between polarizers $\alpha - \beta$ equals the angle between Hilbert space measurement bases, this result immediately generalizes to conservation per NPRF of whatever the outcomes represent when like outcomes entail conservation. [We will expand on the general conservation principle below.]    
Consequently, the CHSH quantity that obtains using correlations derived by demanding strict (each measurement pair) conservation for like settings (same reference frame, defining the entangled state) and average conservation for unlike settings (different reference frames) is exactly that of quantum mechanics. That explains the Tsirelson bound per conservation of binary information over a spatiotemporal ensemble, i.e., conservation per NPRF. Accordingly, expecting the Bell inequality to be satisfied per classical probability theory means selectively abandoning the conservation principle proposed here. 

We can now show how superquantum correlations in accord with the no-signaling condition that can exceed the Tsirelson bound, violate our spacetime symmetry group constraint. We already know that superquantum correlations must violate this constraint, since the constraint yields quantum correlations and superquantum correlations exceed quantum correlations. This merely serves as an example for clarity. The Popescu-Rohrlich (PR) correlations \citep{PR1994}
\begin{equation}
\begin{split}
&p(1,1 \mid a,b) = p(-1,-1 \mid a, b)=\frac{1}{2}\\
&p(1,1 \mid a,b^\prime) = p(-1,-1 \mid a, b^\prime)=\frac{1}{2}\\
&p(1,1 \mid a^\prime,b) = p(-1,-1 \mid a^\prime, b)=\frac{1}{2}\\
&p(1,-1 \mid a^\prime,b^\prime) = p(-1,1 \mid a^\prime, b^\prime)=\frac{1}{2} \label{PRcorr}
\end{split}
\end{equation}
produce a value of 4 for Eq. (\ref{CHSH1}), the largest of any no-signaling possibilities. In order to explicitly relate quantum and superquantum correlations, we bring Eq. (\ref{PRcorr}) to bear on our spin singlet and Mermin photon states. Again, we will focus the discussion on the spin singlet state and allude to the obvious manner by which the analysis carries over to the Mermin photon state.

The last PR correlation certainly makes sense if $a^\prime = b^\prime$, i.e., the total anti-correlation implying conservation of angular momentum, so let us start there. The third PR correlation makes sense for $b = \pi + b^\prime$, where we have conservation of angular momentum with Bob having flipped his coordinate directions. Likewise, then, the second PR correlation makes sense for $a = \pi + a^\prime$, where we have conservation of angular momentum with Alice having flipped her coordinate directions. All of this is perfectly self-consistent with our constraint, since $a^\prime$ and $b^\prime$ are arbitrary per rotational invariance in the spatial plane orthogonal to the source collimation. But now, the first PR correlation is totally at odds with conservation of angular momentum. Both Alice and Bob simply flip their coordinate directions, so we should be right back to the fourth PR correlation with $a^\prime \rightarrow a$ and $b^\prime \rightarrow b$. Instead, the PR correlations say that we have total correlation (maximal violation of conservation of angular momentum) rather than total anti-correlation per conservation of angular momentum, which violates every other observation. In other words, the set of PR observations violates conservation of angular momentum in a maximal sense. To obtain the corresponding argument for angular momentum conservation per the correlated outcomes of the Mermin photon state, simply start with the first PR correlation and show the last PR correlation maximally violates angular momentum conservation. 

To show the spectrum on which superquantum correlations violate our constraint in this context, replace the first PR correlation with
\begin{equation}
\begin{split}
&p(1,1 \mid a,b) = C \\
&p(-1,-1 \mid a, b) = D \\
&p(1,-1 \mid a,b) = E \\
&p(-1,1 \mid a, b) = F \\ \label{PRcorrMod}
\end{split}
\end{equation}
The no-signaling condition Eq. (\ref{nosig}) in conjunction with the second and third PR correlations gives $C = D$ and $E = F$. That in conjunction with normalization $C + D + E + F =1$ and p(anti-correlation) + p(correlation) = 1 means total anti-correlation is the conservation of angular momentum per the quantum case while total correlation is the max violation of conservation of angular momentum per the PR case. Thus, we have a spectrum of superquantum correlations all violating conservation of angular momentum. The take-home message is that if the correlation is stronger than that of quantum mechanics, it violates conservation of angular momentum per NPRF.

Using Eq. (\ref{PRcorrMod}) with the second, third, and fourth PR correlations we obtain a CHSH quantity of 
\begin{equation}
3 + C + D - E - F \label{PRModCHSH}
\end{equation}
As we pointed out above, the PR correlations Eq. (\ref{PRcorr}) have $C = D = 1/2$ and $E = F = 0$ giving a CHSH quantity of 4. And, the angular-momentum-conserving quantum correlations are $C = D = 0$ and $E = F = 1/2$ giving a CHSH quantity of 2. Thus in this case, the superquantum correlations violate conservation of angular momentum when the quantum correlation is below the Tsirelson bound.  

This conclusion also follows from the Mermin photon state by replacing the last PR correlation in analogous fashion, again with $\theta = \pi$. We let $a \rightarrow a^\prime$ and $b \rightarrow b^\prime$ in Eq. (\ref{PRcorrMod}) to replace the fourth PR correlation and the CHSH quantity becomes
\begin{equation}
3 - C - D  + E + F \label{PRModCHSHMermin}
\end{equation}
Now the max violation of conservation of angular momentum occurs for $E = F = 1/2$ (PR case) and total conservation of angular momentum occurs for $C = D = 1/2$ (quantum case). Again, the quantum correlation gives a CHSH quantity of 2 for this case, satisfying the Bell inequality. Thus, again, the superquantum correlations violate conservation of angular momentum when the quantum correlation is below the Tsirelson bound. What is important to see is that correlations stronger than those of quantum mechanics violate conservation of angular momentum. Since we have not observed such violations of conservation of angular momentum (CHSH quantity in excess of the the quantum prediction \citep{poh}), we can rule out superquantum correlations on empirical grounds (Figure \ref{Summary}).

This spin singlet and Mermin photon state analysis generalizes to any measurement associated with any of the four Bell basis states 
\begin{equation}
\begin{split}
|\psi_-\rangle = &\frac{1}{\sqrt{2}} \left(\mid +1-1 \rangle - \mid -1+1 \rangle \right)\\
|\psi_+\rangle = &\frac{1}{\sqrt{2}} \left(\mid +1-1 \rangle + \mid -1+1 \rangle \right)\\
|\phi_+\rangle = &\frac{1}{\sqrt{2}} \left(\mid +1+1 \rangle + \mid -1-1 \rangle \right)\\
|\phi_-\rangle = &\frac{1}{\sqrt{2}} \left(\mid +1+1 \rangle - \mid -1-1 \rangle \right)\\ \label{bellstates}
\end{split}
\end{equation}
(in the $\sigma_z$ eigenbasis, say). The eigenvalues for any $2\times 2$ Hermitian matrix can be written +1 and --1, so whatever Alice and Bob are measuring it gives outcomes of +1 or --1 (NPRF). $|\psi_-\rangle$ is invariant under all three SU(2) transformations $e^{i\theta\sigma_i}$, where $\sigma_i$ are the Pauli spin matrices with $i = \{x,y,z\}$. $|\psi_+\rangle$ is invariant under $e^{i\theta\sigma_z}$, $|\phi_-\rangle$ is invariant under $e^{i\theta\sigma_x}$, and $|\phi_+\rangle$ is invariant under $e^{i\theta\sigma_y}$ \citep{MerminChallenge}. The binary outcomes represent bi-directionality in the plane of symmetry as with spin-$\frac{1}{2}$ particles, or they represent axial duality perpendicular to the plane of symmetry as with spin-$1$ particles. So, whatever the various measurement settings represent, Bob and Alice always measure unlike results in like settings for the first state (``unlike state'') and like results for like settings for the last three states (``like states'') in the plane of symmetry\footnote{$|\psi_+\rangle$ is $i|\phi_+\rangle$ in the $\sigma_y$ eigenbasis and $-|\phi_-\rangle$ in the $\sigma_x$ eigenbasis, so it gives like results for like settings in the $xy$ plane of symmetry.}, as required for explicit conservation of binary information, e.g., on-on(ness) or on-off(ness) or yes-no(ness), when Alice and Bob make the same measurement. Further, Bob and Alice measure +1 and --1 with equal frequency regardless of their settings (NPRF), so Alice's results can be split into two equal sets of +1 and --1 outcomes. For her +1 results, conservation dictates Bob's outcomes will average $[\mbox{cos}^2\theta - \mbox{sin}^2\theta]$ for the three like states and $[\mbox{sin}^2\theta - \mbox{cos}^2\theta]$ for the unlike state where $\theta$ is now the angle between eigenbases in Hilbert space representing whatever the relative difference in settings means in spacetime. For Alice's --1 results, the two equations are flipped, so we have correlations of $\pm\cos{2\theta}$ for like and unlike states, respectively, which give the Tsirelson bound. Of course, Bob can make the same claim about Alice's outcomes satisfying conservation on average with his +1 and --1 outcomes. 

For spin-$\frac{1}{2}$ particles, $\theta$ between bases in Hilbert space is half the angle $\alpha - \beta$ in real space due to bi-directionality in the plane of symmetry. And, for spin-$1$ particles, $\alpha - \beta$ in real space is equal to the angle $\theta$ between bases in Hilbert space due to axial duality perpendicular to the plane of symmetry. Thus, the derivations of the quantum correlations and quantum states using our constraint and NPRF proper for the spin singlet and Mermin photon cases immediately generalize to any Bell basis state analysis in Hilbert space. Consequently, our constraint in a very general sense is conservation per NPRF.

Classical correlations violate the constraint by not reaching the upper limit of the quantum correlations (Tsirelson bound) as usual, and superquantum correlations violate the constraint by exceeding the quantum correlations. Specifically, the probability of measuring unlike results for the unlike state is $\mbox{cos}^2\theta$ and the probability of measuring like results is $\mbox{sin}^2\theta$ (generalized spin singlet case). Similarly, the probability of measuring like results for the like states is $\mbox{cos}^2\theta$ and the probability of measuring unlike results is $\mbox{sin}^2\theta$ (generalized Mermin photon case). 

\begin{figure}
\begin{center}
\includegraphics [height = 40mm]{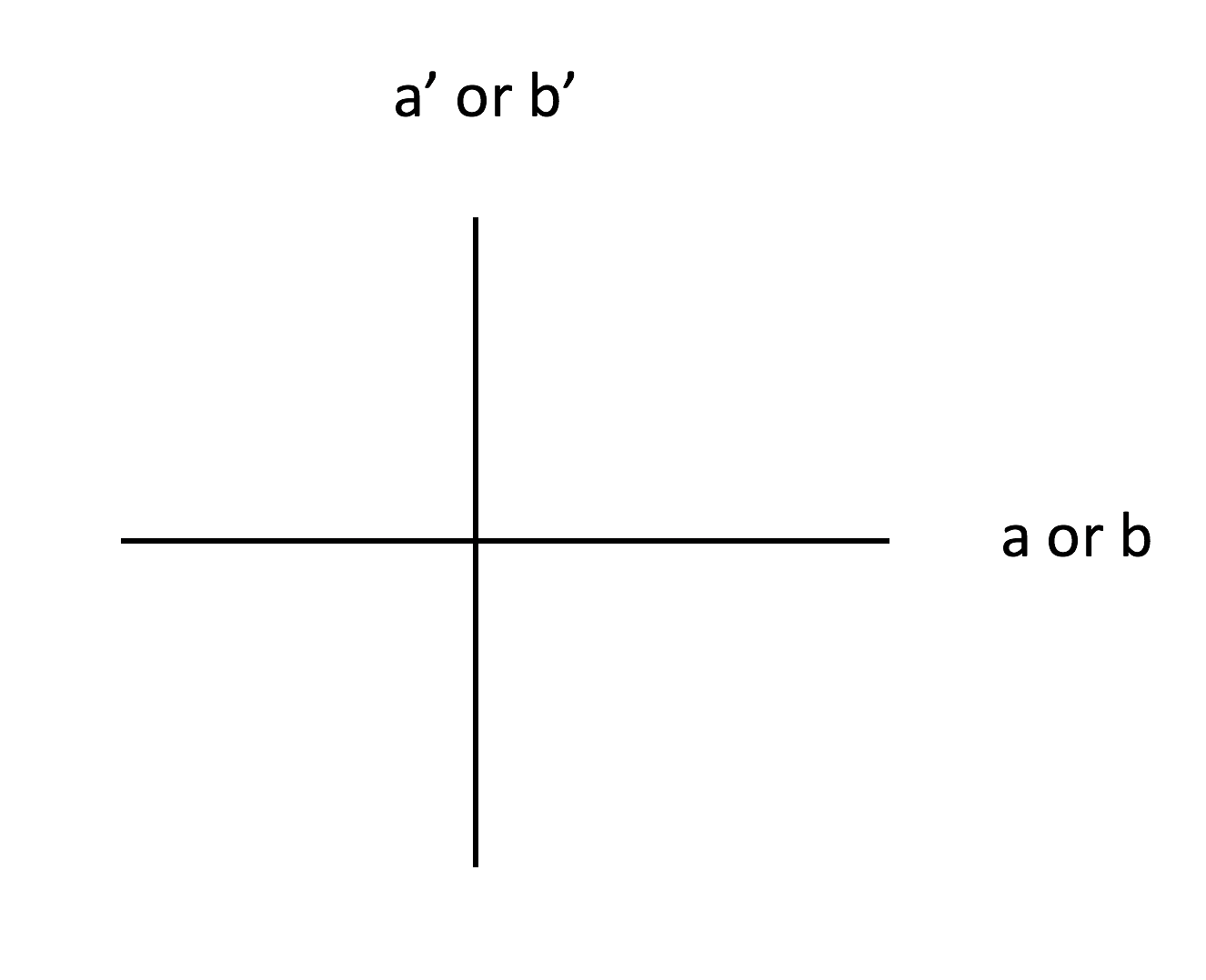}  \caption{Relative eigenbases configuration for the unlike state of Eq. (\ref{bellstates}) satisfying the last three PR correlations.} \label{UnlikeBell}
\end{center}
\end{figure}

Start with the unlike state. The last PR correlation says that $a^\prime$ and $b^\prime$ must be parallel ($\mbox{cos}^2\theta = 1$, Figure \ref{UnlikeBell}). The third PR correlation says $a^\prime$ and $b$ must be perpendicular ($\mbox{sin}^2\theta = 1$, Figure \ref{UnlikeBell}). And, the second PR correlation says $a$ and $b^\prime$ must be perpendicular ($\mbox{sin}^2\theta = 1$, Figure \ref{UnlikeBell}). Thus, these three PR correlations in total say $a^\prime/b^\prime$ is perpendicular to $a/b$ (Figure \ref{UnlikeBell}). So, we need the first PR correlation to say $a$ and $b$ are parallel, but of course it says that $a$ and $b$ must also be perpendicular ($\mbox{sin}^2\theta = 1$). Therefore, the PR correlations violate our constraint in a maximal fashion for the unlike state.

\begin{figure}
\begin{center}
\includegraphics [height = 40mm]{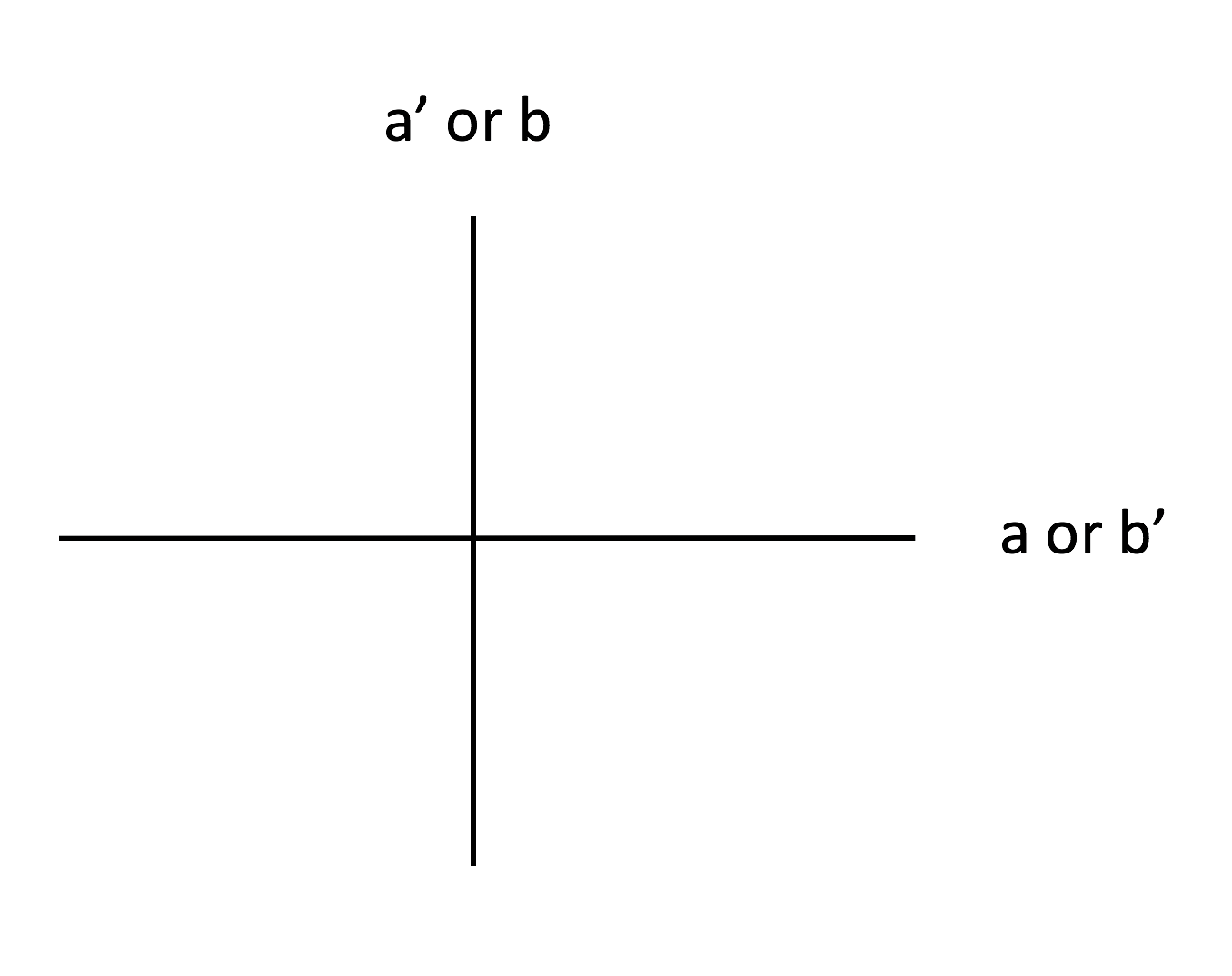}  \caption{Relative eigenbases configuration for the like states of Eq. (\ref{bellstates}) satisfying the last three PR correlations.} \label{LikeBell}
\end{center}
\end{figure}

Now for the like states. The last PR correlation says that $a^\prime$ and $b^\prime$ must be perpendicular ($\mbox{sin}^2\theta = 1$, Figure \ref{LikeBell}). The third PR correlation says $a^\prime$ and $b$ must be parallel ($\mbox{cos}^2\theta = 1$, Figure \ref{LikeBell}). And, the second PR correlation says $a$ and $b^\prime$ must be parallel ($\mbox{cos}^2\theta = 1$, Figure \ref{LikeBell}). Thus, these three PR correlations in total say  $a^\prime/b$ is perpendicular to $a/b^\prime$ (Figure \ref{LikeBell}). So, we need the first PR correlation to say $a$ and $b$ are perpendicular, but of course it says $a$ and $b$ must also be parallel ($\mbox{cos}^2\theta = 1$). Therefore, the PR correlations also violate our constraint in a maximal fashion for the like states.

Again, replacing the first PR correlation with Equation (\ref{PRcorrMod}) and using the no-signaling condition Equation (\ref{nosig}) in conjunction with the second and third PR correlations gives $C = D$ and $E = F$. The CHSH quantity is $3 + C + D - E - F$ for both the unlike and like states. We just showed that for both like and unlike states we need $C = D = 0$ and $E = F = 1/2$ to satisfy conservation per NPRF, but the PR correlations are just the opposite, $C = D = 1/2$ and $E = F = 0$. This shows how superquantum correlations violate our constraint in a very general sense.

\section{\label{disc}Discussion}
Given the widely recognized fundamental importance of conservation principles following from the spacetime symmetry group, this spin singlet state and Mermin photon state analysis suggests that our constraint-based approach provides a ``motivated principle'' requested by Cuffaro \citep{cuffaro1}. And, just as the light postulate of special relativity is in accord with NPRF, the Bell basis states producing the Tsirelson bound are also in accord with NPRF. Alice and Bob each measure +1 and --1 with equal frequency for all measurement settings, they confirm conservation in all trials where their measurement settings (reference frames) are the same, and Alice(Bob) can argue that Bob(Alice) averages less than 1 to satisfy conservation in her(his) choice of measurement setting when the settings (reference frames) are not the same (Figures \ref{4Dpattern} \& \ref{AvgView}).

Thus, our constraint-based/principle answer to Bub's question, ``Why the Tsirelson bound?'' \citep{bub2004,bub2012,bubbook}, the QIT counterpart to Wheeler's ``Really Big Question'', ``Why the quantum?'', can be summed up in ``one clear, simple sentence'' 
\begin{quote}
The Tsirelson bound obtains because of conservation per no preferred reference frame.
\end{quote}
in accord with Fuchs' desideratum (Figure \ref{SRvQM}).

The bottom line is that a compelling constraint (who would argue with conservation per NPRF?) answers ``Why the Tsirelson bound?'' without a corresponding `dynamical/causal influence' or hidden variables to account for the results on a trial-by-trial basis. By accepting the constraint-based explanation as fundamental, the lack of a compelling, consensus dynamical interpretation is not a problem. This is just one of many mysteries in physics created by dynamical thinking and resolved by constraint-based thinking \citep{ourbook}.

Since the Tsirelson bound follows more generally from the mathematical form of any Bell basis state, regardless of what that state represents physically, one could argue that the Tsirelson bound results more generally or more fundamentally from a conservation of binary information. That is, from an information-theoretic perspective, the Tsirelson bound results from a frame independent conservation of on-on(ness) or on-off(ness) or yes-no(ness), etc. However, this conservation of binary information is not information exchanged from Alice(Bob) to Bob(Alice) as in information causality, but information distributed among both Alice and Bob via a spatiotemporal ensemble. Nonetheless, the constraint-based approach certainly negates the need for dynamical interpretation, which is a QIT desideratum. 

One of the many motivations for the QIT approach is the suspicion that realism about the formal machinery of ordinary textbook quantum mechanics is not warranted. For example, the measurement problem is driven by taking seriously the unitary evolution of quantum states in Hilbert space. Likewise quantum entanglement is thought to require nonlocal causal influences, inexplicable non-separability/holism, backwards causation, etc., precisely because it is assumed there must be some dynamical or causal explanation of some sort. However, as we have shown herein, without invoking mere instrumentalism or deflationary tactics, one can explain quantum entanglement and the Tsirelson bound without any of that dynamical baggage and without realism about Hilbert space. And just as the QIT community hypothesized, the explanation is a constraint-based/principle one.

Therefore, we believe our constraint-based answer to ``Why the Tsirelson bound?'' not only justifies the spirit of the QIT constraint-based approach, but provides a physical model of the quantum in spacetime per the interpretation-project (Figures \ref{LikeOrient}, \ref{4Dpattern}, \& \ref{AvgView}), albeit not a dynamical model. As things stand now, there is no obvious connection between the interpretation-project and the QIT-project \citep{timpsonbook}. In this regard, keep in mind that the postulates of special relativity are about the physical world in spacetime; thus, in keeping with this analogy, QIT must eventually make such correspondence to reach their lofty goals and escape the clever, but inherent instrumentalism of ordinary quantum. Our explanation of the Tsirelson bound precisely addresses the need for QIT to make correspondence with phenomena in spacetime. 

The resistance to Einstein's light postulate was immense because he posited something as axiomatic that most people wanted to have explained. We understand the resistance to our constraint, conservation per NPRF, will be at least as fierce even though it satisfies QIT's stated desideratum, i.e., they wanted to answer ``Why the Tsireslon bound?'' per postulates like those of special relativity. So ironically, some in QIT may not be satisfied with our constraint-based/principle explanation of the Tsirelson bound precisely because it posits conservation per NPRF without further explanation; they desire some `extra-physical' explanation for this constraint, e.g., information causality. In the parlance of metaphysics, the complaint is that we have not ruled out some ``possible world'' in which superquantum correlations exist, we have only ruled them out on empirical grounds for our world. On this way of thinking, superquantum correlations can only be ruled out via mathematical or logical necessity from some extra-physics basis. 

Herein, we have answered Bub's question within the purview of physics in precise analogy with the light postulate of special relativity in accord with Fuchs' desideratum, i.e., the Tsirelson bound ultimately obtains because of conservation in accord with no preferred reference frame. Accordingly, there is nothing mysterious about the physics we have discovered so far, it points unambiguously to a physical reality governed fundamentally by constraints. Just as the QIT community suspected, mysteries in quantum mechanics arise solely from the misplaced dynamical expectations of its practitioners. Once we recognize the power of constraints to dispel the mysteries of physics, ``we will all say to each other, how could it have been otherwise? How could we have been so stupid for so long?'' \citep{wheeler1986}.

\bibliography{biblio.bib}

\begin{thebibliography}{39}%
\makeatletter
\providecommand \@ifxundefined [1]{%
 \@ifx{#1\undefined}
}%
\providecommand \@ifnum [1]{%
 \ifnum #1\expandafter \@firstoftwo
 \else \expandafter \@secondoftwo
 \fi
}%
\providecommand \@ifx [1]{%
 \ifx #1\expandafter \@firstoftwo
 \else \expandafter \@secondoftwo
 \fi
}%
\providecommand \natexlab [1]{#1}%
\providecommand \enquote  [1]{``#1''}%
\providecommand \bibnamefont  [1]{#1}%
\providecommand \bibfnamefont [1]{#1}%
\providecommand \citenamefont [1]{#1}%
\providecommand \href@noop [0]{\@secondoftwo}%
\providecommand \href [0]{\begingroup \@sanitize@url \@href}%
\providecommand \@href[1]{\@@startlink{#1}\@@href}%
\providecommand \@@href[1]{\endgroup#1\@@endlink}%
\providecommand \@sanitize@url [0]{\catcode `\\12\catcode `\$12\catcode
  `\&12\catcode `\#12\catcode `\^12\catcode `\_12\catcode `\%12\relax}%
\providecommand \@@startlink[1]{}%
\providecommand \@@endlink[0]{}%
\providecommand \url  [0]{\begingroup\@sanitize@url \@url }%
\providecommand \@url [1]{\endgroup\@href {#1}{\urlprefix }}%
\providecommand \urlprefix  [0]{URL }%
\providecommand \Eprint [0]{\href }%
\providecommand \doibase [0]{http://dx.doi.org/}%
\providecommand \selectlanguage [0]{\@gobble}%
\providecommand \bibinfo  [0]{\@secondoftwo}%
\providecommand \bibfield  [0]{\@secondoftwo}%
\providecommand \translation [1]{[#1]}%
\providecommand \BibitemOpen [0]{}%
\providecommand \bibitemStop [0]{}%
\providecommand \bibitemNoStop [0]{.\EOS\space}%
\providecommand \EOS [0]{\spacefactor3000\relax}%
\providecommand \BibitemShut  [1]{\csname bibitem#1\endcsname}%
\let\auto@bib@innerbib\@empty
\bibitem [{\citenamefont {Wheeler}(1986)}]{wheeler1986}%
  \BibitemOpen
  \bibfield  {author} {\bibinfo {author} {\bibfnamefont {John}\ \bibnamefont
  {Wheeler}},\ }\bibfield  {title} {\enquote {\bibinfo {title} {How come the
  quantum?}}\ }\href@noop {} {\bibfield  {journal} {\bibinfo  {journal} {New
  Techniques and Ideas in Quantum Measurement Theory}\ }\textbf {\bibinfo
  {volume} {480}},\ \bibinfo {pages} {304--316} (\bibinfo {year} {1986})},\
  \bibinfo {note}
  {\url{https://nyaspubs.onlinelibrary.wiley.com/doi/abs/10.1111/j.1749-6632.1986.tb12434.x}}\BibitemShut
  {NoStop}%
\bibitem [{\citenamefont {Hardy}(2016)}]{hardy2016}%
  \BibitemOpen
  \bibfield  {author} {\bibinfo {author} {\bibfnamefont {Lucien}\ \bibnamefont
  {Hardy}},\ }\bibfield  {title} {\enquote {\bibinfo {title} {Reconstructing
  quantum theory},}\ }in\ \href@noop {} {\emph {\bibinfo {booktitle} {Quantum
  Theory: Informational Foundations and Foils}}},\ \bibinfo {editor} {edited
  by\ \bibinfo {editor} {\bibfnamefont {G.}~\bibnamefont {Chiribella}}\ and\
  \bibinfo {editor} {\bibfnamefont {R.}~\bibnamefont {Spekkens}}}\ (\bibinfo
  {publisher} {Springer},\ \bibinfo {address} {Dordrecht},\ \bibinfo {year}
  {2016})\ pp.\ \bibinfo {pages} {223--248},\ \bibinfo {note}
  {\url{https://arxiv.org/abs/1303.1538}}\BibitemShut {NoStop}%
\bibitem [{\citenamefont {Fuchs}\ and\ \citenamefont {Stacey}(2016)}]{fuchs1}%
  \BibitemOpen
  \bibfield  {author} {\bibinfo {author} {\bibfnamefont {C.A.}\ \bibnamefont
  {Fuchs}}\ and\ \bibinfo {author} {\bibfnamefont {B.C.}\ \bibnamefont
  {Stacey}},\ }\bibfield  {title} {\enquote {\bibinfo {title} {Some negative
  remarks on operational approaches to quantum theory},}\ }in\ \href@noop {}
  {\emph {\bibinfo {booktitle} {Quantum Theory: Informational Foundations and
  Foils}}},\ \bibinfo {editor} {edited by\ \bibinfo {editor} {\bibfnamefont
  {G.}~\bibnamefont {Chiribella}}\ and\ \bibinfo {editor} {\bibfnamefont
  {R.}~\bibnamefont {Spekkens}}}\ (\bibinfo  {publisher} {Springer},\ \bibinfo
  {address} {Dordrecht},\ \bibinfo {year} {2016})\ pp.\ \bibinfo {pages}
  {283--305}\BibitemShut {NoStop}%
\bibitem [{\citenamefont {Chiribella}\ and\ \citenamefont
  {Spekkens}(2016)}]{chiribella1}%
  \BibitemOpen
  \bibfield  {author} {\bibinfo {author} {\bibfnamefont {G.}~\bibnamefont
  {Chiribella}}\ and\ \bibinfo {author} {\bibfnamefont {R.}~\bibnamefont
  {Spekkens}},\ }\bibfield  {title} {\enquote {\bibinfo {title}
  {Introduction},}\ }in\ \href@noop {} {\emph {\bibinfo {booktitle} {Quantum
  Theory: Informational Foundations and Foils}}},\ \bibinfo {editor} {edited
  by\ \bibinfo {editor} {\bibfnamefont {G.}~\bibnamefont {Chiribella}}\ and\
  \bibinfo {editor} {\bibfnamefont {R.}~\bibnamefont {Spekkens}}}\ (\bibinfo
  {publisher} {Springer},\ \bibinfo {address} {Dordrecht},\ \bibinfo {year}
  {2016})\ pp.\ \bibinfo {pages} {1--18}\BibitemShut {NoStop}%
\bibitem [{\citenamefont {Wheeler}(1989)}]{wheeler1989}%
  \BibitemOpen
  \bibfield  {author} {\bibinfo {author} {\bibfnamefont {John~Archibald}\
  \bibnamefont {Wheeler}},\ }\bibfield  {title} {\enquote {\bibinfo {title}
  {Information, physics, quantum: The search for links},}\ }in\ \href@noop {}
  {\emph {\bibinfo {booktitle} {Proceedings III International Symposium on
  Foundations of Quantum Mechanics}}},\ \bibinfo {editor} {edited by\ \bibinfo
  {editor} {\bibfnamefont {Shun-ichi~Kobayashi}\ \bibnamefont {et~al.}}}\
  (\bibinfo {year} {1989})\ pp.\ \bibinfo {pages} {354--358}\BibitemShut
  {NoStop}%
\bibitem [{\citenamefont {Barrow}\ \emph {et~al.}(2004)\citenamefont {Barrow},
  \citenamefont {Davies},\ and\ \citenamefont {Charles
  L.~Harper}}]{wheeler2004}%
  \BibitemOpen
  \bibinfo {editor} {\bibfnamefont {John~D.}\ \bibnamefont {Barrow}}, \bibinfo
  {editor} {\bibfnamefont {Paul C.~W.}\ \bibnamefont {Davies}}, \ and\ \bibinfo
  {editor} {\bibfnamefont {Jr.}\ \bibnamefont {Charles L.~Harper}},\ eds.,\
  \href@noop {} {\emph {\bibinfo {title} {Science and Ultimate Reality: Quantum
  Theory, Cosmology, and Complexity}}}\ (\bibinfo  {publisher} {Cambridge
  university Press},\ \bibinfo {address} {New York},\ \bibinfo {year}
  {2004})\BibitemShut {NoStop}%
\bibitem [{\citenamefont {Bub}(2004)}]{bub2004}%
  \BibitemOpen
  \bibfield  {author} {\bibinfo {author} {\bibfnamefont {Jeffrey}\ \bibnamefont
  {Bub}},\ }\bibfield  {title} {\enquote {\bibinfo {title} {Why the quantum?}}\
  }\href@noop {} {\bibfield  {journal} {\bibinfo  {journal} {Studies in History
  and Philosophy of Modern Physics}\ }\textbf {\bibinfo {volume} {35B}},\
  \bibinfo {pages} {241--266} (\bibinfo {year} {2004})},\ \bibinfo {note}
  {\url{https://arxiv.org/abs/quant-ph/0402149}}\BibitemShut {NoStop}%
\bibitem [{\citenamefont {Bub}(2012)}]{bub2012}%
  \BibitemOpen
  \bibfield  {author} {\bibinfo {author} {\bibfnamefont {Jeffrey}\ \bibnamefont
  {Bub}},\ }\bibfield  {title} {\enquote {\bibinfo {title} {Why the {T}sirelson
  bound?}}\ }in\ \href@noop {} {\emph {\bibinfo {booktitle} {The Probable and
  the Improbable: The Meaning and Role of Probability in Physics}}},\ \bibinfo
  {editor} {edited by\ \bibinfo {editor} {\bibfnamefont {Meir}\ \bibnamefont
  {Hemmo}}\ and\ \bibinfo {editor} {\bibfnamefont {Yemima}\ \bibnamefont
  {Ben-Menahem}}}\ (\bibinfo  {publisher} {Springer},\ \bibinfo {address}
  {Dordrecht},\ \bibinfo {year} {2012})\ pp.\ \bibinfo {pages} {167--185},\
  \bibinfo {note} {\url{https://arxiv.org/abs/1208.3744}}\BibitemShut {NoStop}%
\bibitem [{\citenamefont {Bub}(2016)}]{bubbook}%
  \BibitemOpen
  \bibfield  {author} {\bibinfo {author} {\bibfnamefont {Jeffrey}\ \bibnamefont
  {Bub}},\ }\href@noop {} {\emph {\bibinfo {title} {Bananaworld: Quantum
  Mechanics for Primates}}}\ (\bibinfo  {publisher} {Oxford University Press},\
  \bibinfo {address} {Oxford, UK},\ \bibinfo {year} {2016})\BibitemShut
  {NoStop}%
\bibitem [{\citenamefont {Clauser}\ \emph {et~al.}(1969)\citenamefont
  {Clauser}, \citenamefont {Horne}, \citenamefont {Shimony},\ and\
  \citenamefont {Holt}}]{CHSH}%
  \BibitemOpen
  \bibfield  {author} {\bibinfo {author} {\bibfnamefont {John~F.}\ \bibnamefont
  {Clauser}}, \bibinfo {author} {\bibfnamefont {Michael~A.}\ \bibnamefont
  {Horne}}, \bibinfo {author} {\bibfnamefont {Abner}\ \bibnamefont {Shimony}},
  \ and\ \bibinfo {author} {\bibfnamefont {Richard~A.}\ \bibnamefont {Holt}},\
  }\bibfield  {title} {\enquote {\bibinfo {title} {Proposed experiment to test
  local hidden-variable theories},}\ }\href {\doibase
  10.1103/PhysRevLett.23.880} {\bibfield  {journal} {\bibinfo  {journal} {Phys.
  Rev. Lett.}\ }\textbf {\bibinfo {volume} {23}},\ \bibinfo {pages} {880--884}
  (\bibinfo {year} {1969})},\ \bibinfo {note}
  {\url{https://link.aps.org/doi/10.1103/PhysRevLett.23.880}}\BibitemShut
  {NoStop}%
\bibitem [{\citenamefont {Bell}(1964)}]{bell}%
  \BibitemOpen
  \bibfield  {author} {\bibinfo {author} {\bibfnamefont {J.}~\bibnamefont
  {Bell}},\ }\bibfield  {title} {\enquote {\bibinfo {title} {On the
  einstein-podolsky-rosen paradox},}\ }\href@noop {} {\bibfield  {journal}
  {\bibinfo  {journal} {Physics}\ }\textbf {\bibinfo {volume} {1}},\ \bibinfo
  {pages} {195--200} (\bibinfo {year} {1964})}\BibitemShut {NoStop}%
\bibitem [{\citenamefont {Cirel'son}(1980)}]{cirelson1980}%
  \BibitemOpen
  \bibfield  {author} {\bibinfo {author} {\bibfnamefont {B.S.}\ \bibnamefont
  {Cirel'son}},\ }\bibfield  {title} {\enquote {\bibinfo {title} {Quantum
  generalizations of bell's inequality},}\ }\href@noop {} {\bibfield  {journal}
  {\bibinfo  {journal} {Letters in Mathematical Physics}\ }\textbf {\bibinfo
  {volume} {4}},\ \bibinfo {pages} {93--100} (\bibinfo {year} {1980})},\
  \bibinfo {note}
  {\url{https://www.tau.ac.il/~tsirel/download/qbell80.pdf}}\BibitemShut
  {NoStop}%
\bibitem [{\citenamefont {Popescu}\ and\ \citenamefont
  {Rohrlich}(1994)}]{PR1994}%
  \BibitemOpen
  \bibfield  {author} {\bibinfo {author} {\bibfnamefont {S.}~\bibnamefont
  {Popescu}}\ and\ \bibinfo {author} {\bibfnamefont {D.}~\bibnamefont
  {Rohrlich}},\ }\bibfield  {title} {\enquote {\bibinfo {title} {Quantum
  nonlocality as an axiom},}\ }\href@noop {} {\bibfield  {journal} {\bibinfo
  {journal} {Foundations of Physics}\ }\textbf {\bibinfo {volume} {24}},\
  \bibinfo {pages} {379--385} (\bibinfo {year} {1994})},\ \bibinfo {note}
  {\url{https://arxiv.org/abs/quant-ph/9508009v1}}\BibitemShut {NoStop}%
\bibitem [{\citenamefont {Buhrman}\ and\ \citenamefont
  {Massar}(2005)}]{buhrman}%
  \BibitemOpen
  \bibfield  {author} {\bibinfo {author} {\bibfnamefont {H.}~\bibnamefont
  {Buhrman}}\ and\ \bibinfo {author} {\bibfnamefont {S.}~\bibnamefont
  {Massar}},\ }\bibfield  {title} {\enquote {\bibinfo {title} {Causality and
  tsirelson's bounds},}\ }\href {\doibase 10.1103/PhysRevA.72.052103}
  {\bibfield  {journal} {\bibinfo  {journal} {Phys. Rev. A}\ }\textbf {\bibinfo
  {volume} {72}},\ \bibinfo {pages} {052103} (\bibinfo {year} {2005})},\
  \bibinfo {note}
  {\url{https://link.aps.org/doi/10.1103/PhysRevA.72.052103}}\BibitemShut
  {NoStop}%
\bibitem [{\citenamefont {Pawlowski}\ \emph {et~al.}(2009)\citenamefont
  {Pawlowski}, \citenamefont {Paterek}, \citenamefont {Kaszlikowski},
  \citenamefont {Scarani}, \citenamefont {Winter},\ and\ \citenamefont
  {Zukowski}}]{pawlowski2009}%
  \BibitemOpen
  \bibfield  {author} {\bibinfo {author} {\bibfnamefont {M.}~\bibnamefont
  {Pawlowski}}, \bibinfo {author} {\bibfnamefont {T.}~\bibnamefont {Paterek}},
  \bibinfo {author} {\bibfnamefont {D.}~\bibnamefont {Kaszlikowski}}, \bibinfo
  {author} {\bibfnamefont {V.}~\bibnamefont {Scarani}}, \bibinfo {author}
  {\bibfnamefont {A.}~\bibnamefont {Winter}}, \ and\ \bibinfo {author}
  {\bibfnamefont {M.}~\bibnamefont {Zukowski}},\ }\bibfield  {title} {\enquote
  {\bibinfo {title} {Information causality as a physical principle},}\
  }\href@noop {} {\bibfield  {journal} {\bibinfo  {journal} {Nature}\ }\textbf
  {\bibinfo {volume} {461}},\ \bibinfo {pages} {1101--1104} (\bibinfo {year}
  {2009})},\ \bibinfo {note}
  {\url{https://arxiv.org/abs/0905.2292}}\BibitemShut {NoStop}%
\bibitem [{\citenamefont {Cuffaro}()}]{cuffaro1}%
  \BibitemOpen
  \bibfield  {author} {\bibinfo {author} {\bibfnamefont {Michael~E.}\
  \bibnamefont {Cuffaro}},\ }\href@noop {} {\enquote {\bibinfo {title}
  {Information causality, the tsirelson bound, and the `being-thus' of
  things},}\ }\bibinfo {note}
  {\url{http://philsci-archive.pitt.edu/14027/1/tbound.pdf}}\BibitemShut
  {NoStop}%
\bibitem [{\citenamefont {Ball}(2017)}]{Wired}%
  \BibitemOpen
  \bibfield  {author} {\bibinfo {author} {\bibfnamefont {P.}~\bibnamefont
  {Ball}},\ }\bibfield  {title} {\enquote {\bibinfo {title} {Physicists want to
  rebuild quantum theory from scratch},}\ }\href@noop {} {\bibfield  {journal}
  {\bibinfo  {journal} {Wired}\ } (\bibinfo {year} {2017})},\ \bibinfo {note}
  {\url{https://www.wired.com/story/physicists-want-to-rebuild-quantum-theory-from-scratch/amp}}\BibitemShut
  {NoStop}%
\bibitem [{\citenamefont {Silberstein}\ and\ \citenamefont
  {Stuckey}(2019)}]{silber2019}%
  \BibitemOpen
  \bibfield  {author} {\bibinfo {author} {\bibfnamefont {M.}~\bibnamefont
  {Silberstein}}\ and\ \bibinfo {author} {\bibfnamefont {W.M.}\ \bibnamefont
  {Stuckey}},\ }\href@noop {} {\enquote {\bibinfo {title} {Quantum mechanics
  and the consistency of conscious experience},}\ } (\bibinfo {year} {2019}),\
  \bibinfo {note} {\url{https://arxiv.org/abs/1901.10825}}\BibitemShut
  {NoStop}%
\bibitem [{\citenamefont {Silberstein}\ \emph {et~al.}(2018)\citenamefont
  {Silberstein}, \citenamefont {Stuckey},\ and\ \citenamefont
  {McDevitt}}]{ourbook}%
  \BibitemOpen
  \bibfield  {author} {\bibinfo {author} {\bibfnamefont {Michael}\ \bibnamefont
  {Silberstein}}, \bibinfo {author} {\bibfnamefont {W.M.}\ \bibnamefont
  {Stuckey}}, \ and\ \bibinfo {author} {\bibfnamefont {Timothy}\ \bibnamefont
  {McDevitt}},\ }\href@noop {} {\emph {\bibinfo {title} {Beyond the Dynamical
  Universe: Unifying Block Universe Physics and Time as Experienced}}}\
  (\bibinfo  {publisher} {Oxford University Press},\ \bibinfo {address}
  {Oxford, UK},\ \bibinfo {year} {2018})\BibitemShut {NoStop}%
\bibitem [{\citenamefont {Unnikrishnan}(2005)}]{unnik2005}%
  \BibitemOpen
  \bibfield  {author} {\bibinfo {author} {\bibfnamefont {C.~S.}\ \bibnamefont
  {Unnikrishnan}},\ }\bibfield  {title} {\enquote {\bibinfo {title}
  {Correlation functions, bell's inequalities and the fundamental conservation
  laws},}\ }\href@noop {} {\bibfield  {journal} {\bibinfo  {journal}
  {Europhysics Letters}\ }\textbf {\bibinfo {volume} {69}},\ \bibinfo {pages}
  {489--495} (\bibinfo {year} {2005})}\BibitemShut {NoStop}%
\bibitem [{\citenamefont {Cabello}(2002)}]{cabello}%
  \BibitemOpen
  \bibfield  {author} {\bibinfo {author} {\bibfnamefont {Adan}\ \bibnamefont
  {Cabello}},\ }\bibfield  {title} {\enquote {\bibinfo {title} {Violating
  bell's inequality beyond cirel'son's bound},}\ }\href@noop {} {\bibfield
  {journal} {\bibinfo  {journal} {Physical Review Letters}\ }\textbf {\bibinfo
  {volume} {88}},\ \bibinfo {pages} {060403} (\bibinfo {year} {2002})},\
  \bibinfo {note} {\url{https://arxiv.org/abs/quant-ph/0108084}}\BibitemShut
  {NoStop}%
\bibitem [{\citenamefont {Landau}(1987)}]{landau1987}%
  \BibitemOpen
  \bibfield  {author} {\bibinfo {author} {\bibfnamefont {Lawrence~J.}\
  \bibnamefont {Landau}},\ }\bibfield  {title} {\enquote {\bibinfo {title} {On
  the violation of bell's inequality in quantum theory},}\ }\href@noop {}
  {\bibfield  {journal} {\bibinfo  {journal} {Physics Letters A}\ }\textbf
  {\bibinfo {volume} {120}},\ \bibinfo {pages} {54--56} (\bibinfo {year}
  {1987})}\BibitemShut {NoStop}%
\bibitem [{\citenamefont {Khalfin}\ and\ \citenamefont
  {Tsirelson}(1992)}]{khalfin1992}%
  \BibitemOpen
  \bibfield  {author} {\bibinfo {author} {\bibfnamefont {Leonid~A.}\
  \bibnamefont {Khalfin}}\ and\ \bibinfo {author} {\bibfnamefont {Boris~S.}\
  \bibnamefont {Tsirelson}},\ }\bibfield  {title} {\enquote {\bibinfo {title}
  {Quantum/classical correspondence in the light of bell's inequalities},}\
  }\href@noop {} {\bibfield  {journal} {\bibinfo  {journal} {Foundations of
  Physics}\ }\textbf {\bibinfo {volume} {22}},\ \bibinfo {pages} {879--948}
  (\bibinfo {year} {1992})},\ \bibinfo {note}
  {\url{https://www.tau.ac.il/~tsirel/download/quantcl.ps}}\BibitemShut
  {NoStop}%
\bibitem [{\citenamefont {Ghirardi}\ and\ \citenamefont
  {Romano}(2012)}]{ghirardi}%
  \BibitemOpen
  \bibfield  {author} {\bibinfo {author} {\bibfnamefont {GianCarlo}\
  \bibnamefont {Ghirardi}}\ and\ \bibinfo {author} {\bibfnamefont {Raffaele}\
  \bibnamefont {Romano}},\ }\bibfield  {title} {\enquote {\bibinfo {title}
  {Arbitrary violation of the tsirelson bound},}\ }\href {\doibase
  10.1103/PhysRevA.86.022116} {\bibfield  {journal} {\bibinfo  {journal} {Phys.
  Rev. A}\ }\textbf {\bibinfo {volume} {86}},\ \bibinfo {pages} {022116}
  (\bibinfo {year} {2012})},\ \bibinfo {note}
  {https://link.aps.org/doi/10.1103/PhysRevA.86.022116}\BibitemShut {NoStop}%
\bibitem [{\citenamefont {Mermin}(1994)}]{mermin1}%
  \BibitemOpen
  \bibfield  {author} {\bibinfo {author} {\bibfnamefont {N.D.}\ \bibnamefont
  {Mermin}},\ }\bibfield  {title} {\enquote {\bibinfo {title} {Quantum
  mysteries refined},}\ }\href@noop {} {\bibfield  {journal} {\bibinfo
  {journal} {American Journal of Physics}\ }\textbf {\bibinfo {volume} {62}},\
  \bibinfo {pages} {880--887} (\bibinfo {year} {1994})}\BibitemShut {NoStop}%
\bibitem [{\citenamefont {Garg}\ and\ \citenamefont {Mermin}(1982)}]{garg}%
  \BibitemOpen
  \bibfield  {author} {\bibinfo {author} {\bibfnamefont {A.}~\bibnamefont
  {Garg}}\ and\ \bibinfo {author} {\bibfnamefont {N.D.}\ \bibnamefont
  {Mermin}},\ }\bibfield  {title} {\enquote {\bibinfo {title} {Bell
  inequalities with a range of violation that does not diminish as the spin
  becomes arbitrarily large},}\ }\href@noop {} {\bibfield  {journal} {\bibinfo
  {journal} {Phys. Rev. Lett.}\ }\textbf {\bibinfo {volume} {49}},\ \bibinfo
  {pages} {901--904} (\bibinfo {year} {1982})}\BibitemShut {NoStop}%
\bibitem [{\citenamefont {Hardy}(2011)}]{hardy2011}%
  \BibitemOpen
  \bibfield  {author} {\bibinfo {author} {\bibfnamefont {Lucien}\ \bibnamefont
  {Hardy}},\ }\href@noop {} {\enquote {\bibinfo {title} {Reformulating and
  reconstructing quantum theory},}\ } (\bibinfo {year} {2011}),\ \bibinfo
  {note} {\url{https://arxiv.org/abs/1104.2066}}\BibitemShut {NoStop}%
\bibitem [{\citenamefont {Bohm}(1952)}]{bohm}%
  \BibitemOpen
  \bibfield  {author} {\bibinfo {author} {\bibfnamefont {David}\ \bibnamefont
  {Bohm}},\ }\href@noop {} {\emph {\bibinfo {title} {Quantum Theory}}}\
  (\bibinfo  {publisher} {Prentice-Hall},\ \bibinfo {address} {New Jersey,
  USA},\ \bibinfo {year} {1952})\BibitemShut {NoStop}%
\bibitem [{\citenamefont {La~Rosa}()}]{larosa}%
  \BibitemOpen
  \bibfield  {author} {\bibinfo {author} {\bibfnamefont {A.}~\bibnamefont
  {La~Rosa}},\ }\href@noop {} {\enquote {\bibinfo {title} {Introduction to
  quantum mechanics, lecture 12, quantum entanglement},}\ }\bibinfo {note}
  {\url{https://www.pdx.edu/nanogroup/sites/www.pdx.edu.nanogroup/files/QUANTUMENTANGLEMENT_18.pdf}}\BibitemShut
  {NoStop}%
\bibitem [{\citenamefont {Hensen}(2015)}]{hensen}%
  \BibitemOpen
  \bibfield  {author} {\bibinfo {author} {\bibfnamefont {B.}~\bibnamefont
  {Hensen}},\ }\href@noop {} {\enquote {\bibinfo {title} {Experimental
  loophole-free violation of a bell inequality using entangled electron spins
  separated by 1.3 km},}\ } (\bibinfo {year} {2015}),\ \bibinfo {note}
  {\url{https://arxiv.org/pdf/1508.05949.pdf}}\BibitemShut {NoStop}%
\bibitem [{\citenamefont {Dehlinger}\ and\ \citenamefont
  {Mitchell}(2002)}]{dehlinger}%
  \BibitemOpen
  \bibfield  {author} {\bibinfo {author} {\bibfnamefont {D.}~\bibnamefont
  {Dehlinger}}\ and\ \bibinfo {author} {\bibfnamefont {M.W.}\ \bibnamefont
  {Mitchell}},\ }\bibfield  {title} {\enquote {\bibinfo {title} {Entangled
  photons, nonlocality, and bell inequalities in the undergraduate
  laboratory},}\ }\href@noop {} {\bibfield  {journal} {\bibinfo  {journal}
  {American Journal of Physics}\ }\textbf {\bibinfo {volume} {70}},\ \bibinfo
  {pages} {903--910} (\bibinfo {year} {2002})}\BibitemShut {NoStop}%
\bibitem [{\citenamefont {Renes}(2015)}]{renes}%
  \BibitemOpen
  \bibfield  {author} {\bibinfo {author} {\bibfnamefont {J.}~\bibnamefont
  {Renes}},\ }\href@noop {} {\enquote {\bibinfo {title} {Quantum information
  theory},}\ } (\bibinfo {year} {2015}),\ \bibinfo {note}
  {\url{http://edu.itp.phys.ethz.ch/hs15/QIT/renes_lecture_notes14.pdf}}\BibitemShut
  {NoStop}%
\bibitem [{\citenamefont {Stuckey}\ \emph {et~al.}(2019)\citenamefont
  {Stuckey}, \citenamefont {McDevitt}, \citenamefont {Silberstein},\ and\
  \citenamefont {Le}}]{MerminChallenge}%
  \BibitemOpen
  \bibfield  {author} {\bibinfo {author} {\bibfnamefont {W.M.}\ \bibnamefont
  {Stuckey}}, \bibinfo {author} {\bibfnamefont {T.}~\bibnamefont {McDevitt}},
  \bibinfo {author} {\bibfnamefont {M.}~\bibnamefont {Silberstein}}, \ and\
  \bibinfo {author} {\bibfnamefont {T.D.}\ \bibnamefont {Le}},\ }\href@noop {}
  {\enquote {\bibinfo {title} {Answering mermin's challenge: Conservation per
  no preferred reference frame},}\ } (\bibinfo {year} {2019}),\ \bibinfo {note}
  {\url{https://arxiv.org/abs/1809.08231}}\BibitemShut {NoStop}%
\bibitem [{\citenamefont {Sinha}\ and\ \citenamefont {Sorkin}(1991)}]{sinha}%
  \BibitemOpen
  \bibfield  {author} {\bibinfo {author} {\bibfnamefont {S.}~\bibnamefont
  {Sinha}}\ and\ \bibinfo {author} {\bibfnamefont {R.D.}\ \bibnamefont
  {Sorkin}},\ }\bibfield  {title} {\enquote {\bibinfo {title} {A
  sum-over-histories account of an epr(b) experiment},}\ }\href@noop {}
  {\bibfield  {journal} {\bibinfo  {journal} {Foundations of Physics Letters}\
  }\textbf {\bibinfo {volume} {4}},\ \bibinfo {pages} {303--335} (\bibinfo
  {year} {1991})}\BibitemShut {NoStop}%
\bibitem [{\citenamefont {Wharton}\ \emph {et~al.}(2011)\citenamefont
  {Wharton}, \citenamefont {Miller},\ and\ \citenamefont {Price}}]{wharton1}%
  \BibitemOpen
  \bibfield  {author} {\bibinfo {author} {\bibfnamefont {K.}~\bibnamefont
  {Wharton}}, \bibinfo {author} {\bibfnamefont {D.}~\bibnamefont {Miller}}, \
  and\ \bibinfo {author} {\bibfnamefont {H.}~\bibnamefont {Price}},\ }\bibfield
   {title} {\enquote {\bibinfo {title} {Action duality: a constructive
  principle for quantum foundations},}\ }\href@noop {} {\bibfield  {journal}
  {\bibinfo  {journal} {Symmetry}\ }\textbf {\bibinfo {volume} {3}},\ \bibinfo
  {pages} {524--540} (\bibinfo {year} {2011})}\BibitemShut {NoStop}%
\bibitem [{\citenamefont {Mermin}(1981)}]{mermin2}%
  \BibitemOpen
  \bibfield  {author} {\bibinfo {author} {\bibfnamefont {N.D.}\ \bibnamefont
  {Mermin}},\ }\bibfield  {title} {\enquote {\bibinfo {title} {Bringing home
  the atomic world: Quantum mysteries for anybody},}\ }\href@noop {} {\bibfield
   {journal} {\bibinfo  {journal} {American Journal of Physics}\ }\textbf
  {\bibinfo {volume} {49}},\ \bibinfo {pages} {940--943} (\bibinfo {year}
  {1981})}\BibitemShut {NoStop}%
\bibitem [{\citenamefont {Boughn}(2017)}]{boughn}%
  \BibitemOpen
  \bibfield  {author} {\bibinfo {author} {\bibfnamefont {S.}~\bibnamefont
  {Boughn}},\ }\href@noop {} {\enquote {\bibinfo {title} {Making sense of
  bell's theorem and quantum nonlocality},}\ } (\bibinfo {year} {2017}),\
  \bibinfo {note} {\url{https://arxiv.org/abs/1703.11003}}\BibitemShut
  {NoStop}%
\bibitem [{\citenamefont {Poh}\ \emph {et~al.}(2015)\citenamefont {Poh},
  \citenamefont {Joshi}, \citenamefont {Cer\`e}, \citenamefont {Cabello},\ and\
  \citenamefont {Kurtsiefer}}]{poh}%
  \BibitemOpen
  \bibfield  {author} {\bibinfo {author} {\bibfnamefont {Hou~Shun}\
  \bibnamefont {Poh}}, \bibinfo {author} {\bibfnamefont {Siddarth~K.}\
  \bibnamefont {Joshi}}, \bibinfo {author} {\bibfnamefont {Alessandro}\
  \bibnamefont {Cer\`e}}, \bibinfo {author} {\bibfnamefont {Ad\'an}\
  \bibnamefont {Cabello}}, \ and\ \bibinfo {author} {\bibfnamefont {Christian}\
  \bibnamefont {Kurtsiefer}},\ }\bibfield  {title} {\enquote {\bibinfo {title}
  {Approaching tsirelson's bound in a photon pair experiment},}\ }\href
  {\doibase 10.1103/PhysRevLett.115.180408} {\bibfield  {journal} {\bibinfo
  {journal} {Phys. Rev. Lett.}\ }\textbf {\bibinfo {volume} {115}},\ \bibinfo
  {pages} {180408} (\bibinfo {year} {2015})},\ \bibinfo {note}
  {\url{https://link.aps.org/doi/10.1103/PhysRevLett.115.180408}}\BibitemShut
  {NoStop}%
\bibitem [{\citenamefont {Timpson}(2013)}]{timpsonbook}%
  \BibitemOpen
  \bibfield  {author} {\bibinfo {author} {\bibfnamefont {Christopher~G.}\
  \bibnamefont {Timpson}},\ }\href@noop {} {\emph {\bibinfo {title} {Quantum
  Information Theory \& the Foundations of Quantum Mechanics}}}\ (\bibinfo
  {publisher} {Oxford University Press},\ \bibinfo {address} {Oxford, UK},\
  \bibinfo {year} {2013})\BibitemShut {NoStop}%
\end{thebibliography}%

\end{document}